\documentclass[12pt]{elsarticle}       
\usepackage{axodraw4j}
\usepackage{graphicx}                                                         
\usepackage{a41}                                                                   
\usepackage{xcolor}                     
\usepackage{pstricks}           
\usepackage[rflt]{floatflt}
\usepackage{float}
\usepackage{hyperref}
\hypersetup{colorlinks=true, linkcolor=blue, 
filecolor=blue, urlcolor=mygreen}
\usepackage{breakurl} 
\usepackage{times}    %
\usepackage{array}
\usepackage[english]{babel}
\usepackage[latin1]{inputenc}
\usepackage[T1]{fontenc}
\usepackage{ae}
\usepackage{url}
\usepackage{amsmath, amsthm, amssymb}

\usepackage{rotating}
\usepackage{graphicx}
\newcounter{mmacnt}
\def\restartmma{\setcounter{mmacnt}{0}}
\restartmma \catcode`|=\active
\def|#1|{\mathrm{#1}}
\catcode`|=12
\newenvironment{mma}{
\par\smallskip
\catcode`|=\active
\parskip=0pt\parindent=0pt 
\small
\def\In##1\\{%
\def\linebreak{\hfill\break\null\qquad}%
\refstepcounter{mmacnt}
\hangindent=2.5em\hangafter=0
\leavevmode
\llap{\tiny\sffamily In[\arabic{mmacnt}]:=\kern.5em}%
\mathversion{bold}\footnotesize$
\displaystyle##1$\normalsize
\mathversion{normal}\par
 }%
\def\Print##1\\{%
\def\linebreak{\hfill\break}%
\hangindent=2.5em\hangafter=0
\leavevmode ##1\par}%
\def\Out##1\\{%
\def\linebreak{$\hfill\break\null\hfill$}%
\kern\abovedisplayskip\par
\hangindent=2.5em\hangafter=0
\leavevmode
\llap{\tiny\sffamily Out[\arabic{mmacnt}]=\kern.5em}
\footnotesize$\displaystyle##1$
\normalsize\hfill\null\par
\kern\belowdisplayskip
}%
\def\Warning##1##2\\{%
\def\linebreak{\hfill\break}%
\hangindent=2.5em\hangafter=0
\leavevmode
{\scriptsize##1 : ##2}\par}%
}{%
\par\smallskip
}

\usepackage{color}
\newenvironment{fshaded}{%
\MakeFramed {\FrameRestore}
}%
{\endMakeFramed}

\def\ps@pprintTitle{%
\let\@oddhead\@empty
\let\@evenhead\@empty
\def\@oddfoot{\reset@font\hfil\thepage\hfil}
\let\@evenfoot\@oddfoot
}
\makeatother
\usepackage{tikz}
\usetikzlibrary{matrix}
\allowdisplaybreaks[4]
\newcommand{\Fh}[2]{\,{}_#1F_#2}
\newcommand{\Fs}[3]{\!\!\left[\begin{array}{c}#1\,;\\#2\,;
\end{array}#3\right]}
\newcommand{\Fz}[3]{\Fs{#1}{#2}{#3}}
\begin{document}              
\begin{frontmatter}           
\title{\Large\textbf{
Scalar $1$-loop Feynman integrals 
as meromorphic functions in 
space-time dimension $d$, $II$:
Special kinematics} }
\author{Khiem Hong Phan}
\ead{phkhiem@hcmus.edu.vn}
\address{
$^{1)}$University of Science 
Ho Chi Minh City,
$227$ Nguyen Van Cu, 
District $5$, HCM City, 
Vietnam \\
$^{2)}$Vietnam National University 
Ho Chi Minh City, 
Linh Trung Ward, Thu Duc District, 
HCM City, Vietnam}
\pagestyle{myheadings}
\markright{}
\begin{abstract}
Based on the method developed in 
[K.~H.~Phan and T.~Riemann, 
Phys.\ Lett.\ B {\bf 791} (2019) 257],
detailed analytic results for scalar 
one-loop two-, three-, four-point 
integrals in general $d$-dimension
are presented in this paper. 
The calculations are considered 
all external kinematic configurations 
and internal mass assignments. 
Analytic formulas are expressed in terms of 
generalized hypergeometric series 
such as Gauss $_2F_1$, Appell $F_1$ 
and Lauricella $F_S$ functions.
\end{abstract}
\end{frontmatter}
\section{Introduction} 
Scalar one-loop integrals in general 
$d$ are important for several reasons. 
In general framework for computing 
two-loop and higher-loop corrections, 
higher-terms in the $\epsilon$-expansion
($\epsilon =2-d/2$) for one-loop integrals 
are necessary for building blocks. 
For example, they are used for building 
counterterms. Furthermore, 
in the evaluations for multi-loop 
Feynman integrals, we may combine
several methods 
in~\cite{Laporta:2001dd,Tarasov:1996br}
to optimize the master integrals. 
As a result, the resulting integrals
may include of one-loop functions in 
arbitrary space-time dimensions. Last 
but not least, scalar one-loop 
functions at $d= 4+2n \pm 2\epsilon$ 
with $n\in\mathbb{N}$ may be taken into 
account in the reduction for tensor 
one-loop Feynman integrals
\cite{Davydychev:1990cq}.

One-loop functions in space-time 
dimensions $d$ have been performed 
in~\cite{Boos:1990rg,
Davydychev:1997wa,
Davydychev:1999mq,Davydychev:2003mv,
Davydychev:2005nf, Anastasiou:1999ui}.
However, not all of the calculations 
have covered at general configurations 
of external momenta and internal masses.
Recently, scalar one-loop three-point 
functions have been performed by applying 
multiple unitarity cuts for Feynman 
diagrams~\cite{Abreu:2015zaa}. 
In Ref.~\cite{Abreu:2015zaa}, analytic results 
have been presented in terms of 
hypergeometric functions in special cases of 
external invariants and internal masses.
In more general cases, 
the results have only presented $\epsilon^0$-terms
in space-time $d=4-2\epsilon$.
The algebraic 
structure of cut Feynman integrals, 
the diagrammatic coaction and its applications 
have proposed in
\cite{Abreu:2017ptx,Abreu:2017enx,Abreu:2017mtm}. 
However, the detailed analytic results for 
one-loop Feynman integrals have not been 
shown in the mentioned papers.
More recently, detailed
analytic results for one-loop three-point
functions which are expressed
in terms of Appell $F_1$ hypergeometric
functions have been reported 
in~\cite{Phan:2019qee}.

A recurrence relation in $d$ 
for Feynman loop 
integrals has proposed 
by Tarasov~\cite{Tarasov:1996br,
Tarasov:2000sf,Fleischer:2003rm}. 
By solving the differential equation in $d$, 
analytic results for scalar one-loop 
integrals up to 
four-points have been expressed
in terms of generalized 
hypergeometric series such as Gauss ${}_2F_1$, 
Appell $F_1$, 
Lauricella-Saran $F_S$ functions. 
In~\cite{Fleischer:2003rm}, boundary terms have 
obtained by applying asymptotic theory of complex 
Laplace-type integrals. These terms are only 
valid in sub-domain of external 
momentum and mass configurations which the theory 
are applicable. Hence, the general solutions
for one-loop integrals in 
arbitrary kinematics have not been 
found~\cite{Fleischer:2003rm},
as pointed out in~\cite{Bluemlein:2015sia,
Bluemlein:2017rbi}. 
General solutions for this
problem have been derived
in~\cite{Phan:2018cnz} which 
proposed 
a different kind of recursion relation 
for one-loop integrals
in comparison with
\cite{Tarasov:1996br, Tarasov:2000sf,
Fleischer:2003rm}.
In the scope of this paper, based on 
the method in Ref.~\cite{Phan:2018cnz}, 
detailed analytic results for scalar 
one-loop two-, 
three- and four-point functions in 
general $d$-dimensions
are presented. 
The calculations are considered 
all external kinematic configurations 
and internal mass assignments. 
Thus, we go beyond the material 
presented in \cite{Phan:2018cnz}.

The layout of the paper is as follows: 
In section $2$, we discuss briefly the method 
for evaluating one-loop integrals. We then 
apply this method for computing scalar 
one-loop two-, three- and four-point functions 
in sections $3, 4, 5$. Conclusions and plans 
for future work are presented in section $6$. 
\section{Method}
In this section, we describe briefly the 
method for evaluating scalar one-loop 
$N$-point Feynman integrals. 
Detailed description for this method 
can be found in 
Ref.~\cite{Phan:2018cnz}. A general 
recursion relation between scalar
one-loop $N$-point 
and $(N$-$1)$-point Feynman integrals is shown 
in this section. From the representation, 
analytic formulas for scalar one-loop $N$-point 
functions can be constructed from basic 
integrals which are scalar one-point integrals. 
For illustrating, analytic expressions for 
scalar one-loop two-, three-, four-point 
functions are derived in detail in next 
sections. 

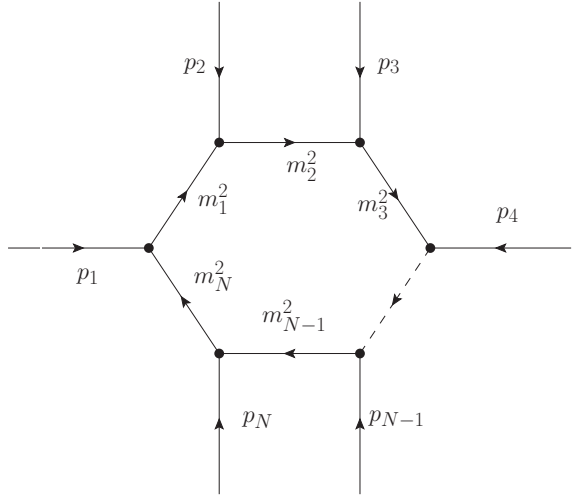
\begin{figure}[ht]
\begin{center}    
\resizebox{0.45\textwidth}{!}{
\fcolorbox{white}{white}{
\begin{picture}(580,508) (330,-70)
\SetWidth{0.1}
\SetColor{Black}
\Line[arrow,arrowpos=0.5,arrowlength=10,
arrowwidth=4,arrowinset=0.2](432,184)(504,292)
\Line[arrow,arrowpos=0.5,arrowlength=10,
arrowwidth=4,arrowinset=0.2](504,292)(648,292)
\Line[arrow,arrowpos=0.5,arrowlength=10,
arrowwidth=4,arrowinset=0.2](648,292)(720,184)
\Line[arrow,arrowpos=0.5,arrowlength=10,
arrowwidth=4,arrowinset=0.2,flip](432,184)(504,76)
\Line[arrow,arrowpos=0.5,arrowlength=10,
arrowwidth=4,arrowinset=0.2,flip](504,76)(648,76)
\Line[dash,dashsize=10,arrow,arrowpos=0.5,
arrowlength=10,arrowwidth=4,
arrowinset=0.2](720,184)(648,76)
\Line[arrow,arrowpos=0.5,arrowlength=10,
arrowwidth=4,arrowinset=0.2](288,184)(432,184)
\Line[arrow,arrowpos=0.5,arrowlength=10,
arrowwidth=4,arrowinset=0.2,flip](504,76)(504,-68)
\Line[arrow,arrowpos=0.5,arrowlength=10,
arrowwidth=4,arrowinset=0.2,flip](648,76)(648,-68)
\Line[arrow,arrowpos=0.5,arrowlength=10,
arrowwidth=4,arrowinset=0.2,flip](720,184)(864,184)
\Line[arrow,arrowpos=0.5,arrowlength=10,
arrowwidth=4,arrowinset=0.2,flip](648,292)(648,436)
\Line[arrow,arrowpos=0.5,arrowlength=10,
arrowwidth=4,arrowinset=0.2,flip](504,292)(504,436)
\Vertex(432,184){5}
\Vertex(504,292){5}
\Vertex(648,292){5}
\Vertex(720,184){5}
\Vertex(648,76){5}
\Vertex(504,76){5}
\Text(660,0)[lb]{\huge{\Black{$p_{N-1}$}}}
\Text(530,0)[lb]{\huge{\Black{$p_N$}}}
\Text(360,148)[lb]{\huge{\Black{$p_1$}}}
\Text(470,364)[lb]{\huge{\Black{$p_2$}}}
\Text(669,364)[lb]{\huge{\Black{$p_3$}}}
\Text(791,211)[lb]{\huge{\Black{$p_4$}}}
\Text(484,220)[lb]{\huge{\Black{$m_1^2$}}}
\Text(575,254)[lb]{\huge{\Black{$m_2^2$}}}
\Text(648,214)[lb]{\huge{\Black{$m_3^2$}}}
\Text(480,140)[lb]{\huge{\Black{$m_N^2$}}}
\Text(550,98)[lb]{\huge{\Black{$m_{N-1}^2$}}}
\end{picture} }}
\end{center}
\caption{\label{Npoints} 
Generic Feynman diagrams 
at one-loop with $N$ external
momenta.}
\end{figure}

The scalar one-loop $N$-point Feynman 
integrals are defined:
\begin{eqnarray}
\label{npoint}
J_{N}(d; \{p_ip_j\}, \{m_i^2\}) 
&=&
\int \dfrac{d^d k}{i \pi^{d/2}}
\dfrac{1}{P_1 P_2\dots P_N}.
\end{eqnarray}
The inverse Feynman propagators 
are given: 
\begin{eqnarray}
P_j=& (k+q_j)^2-m_j^2+i\rho, 
\; \text{for $j=1,2,\cdots, N$}.
\end{eqnarray}
Where $p_j$ ($m_j$) for $j=1,2,\cdots, N$ 
are external momenta (internal masses) 
respectively. The momenta $q_j$ are given:
$q_1 =p_1, q_2 =p_1+p_2, \cdots, 
q_j = \sum_{i=1}^{j}p_i$ and 
$q_N=\sum_{j=1}^{N}p_j=0$ thanks to 
momentum conservation. They are inward as described 
in Fig.~\ref{Npoints}. The term $i\rho$ is Feynman's 
prescription and $d$ is space-time dimension.  
Several cases of physical interests for $d$ are 
$d= 4+2n \pm 2\epsilon$ with $n\in \mathbb{N}$.
In the complex-mass scheme~\cite{Denner:2006ic},
the internal masses take the form of 
$m_j^2 = m_{0j}^2-i m_{0j}\Gamma_{j}$
where $\Gamma_{j}\geqslant 0$ are decay widths of 
unstable particles. 

The Cayley and Gram
determinants~\cite{Fleischer:2003rm}
related to one-loop 
Feynman $N$-point
topologies are defined as follows:
\begin{eqnarray}
Y_N\equiv Y_{12\cdots N}=  \left|
\begin{array}{cccc}
Y_{11}  & Y_{12}  &\ldots & Y_{1N} \\
Y_{12}  & Y_{22}  &\ldots & Y_{2N} \\
\vdots  & \vdots  &\ddots & \vdots \\
Y_{1N}  & Y_{2N}  &\ldots & Y_{NN}
\end{array}
\right|, \hspace{-0.5cm}
\label{Gram}
&& G_{N-1}\equiv 
G_{12\cdots N}=-2^N\; \left|
\begin{array}{cccc}
\! q_1^2 & q_1q_2   &\ldots &q_1q_{N-1} \\
\! q_1q_2  &  q_2^2 &\ldots &q_2q_{N-1} \\
\vdots  & \vdots  &\ddots   &\vdots     \\
\!q_1q_{N-1}  & q_2q_{N-1}  &\ldots& q_{N-1}^2
\end{array}
\right| \nonumber\\
\end{eqnarray}
with $Y_{ij}=-(q_i-q_j)^2+m_i^2+m_j^2$.

In this report, analytic solutions for 
one-loop integrals are expressed 
in terms of generalized hypergeometric 
with arguments given 
by ratios of the above determinants. 
Hence, it is worth to introduce 
the following kinematic variables
\begin{eqnarray}
R_N \equiv R_{12\cdots N}      
&=&-\frac{Y_N }{G_{N-1}}
\quad \text{for} \quad  
G_{N-1}  \neq 0. 
\end{eqnarray}
The kinematics $R_N$ play a role of  
the squared internal masses. In fact, 
when we shift $m_j^2 
\rightarrow m_j^2 - i\rho$, 
one verifies easily that 
$R_N \rightarrow R_N- i\rho$
\cite{Fleischer:2003rm}.

The recursion relation for $J_N$
\cite{Phan:2018cnz} is given
(master equation):
\begin{eqnarray}
\label{JNJN1}
J_{N}(d; \{p_ip_j\}, \{m_i^2\})
&=& -\dfrac{1}{2\pi i} 
\int\limits_{-i\infty}^{+i\infty}ds \; 
\dfrac{\Gamma(-s)\;
\Gamma(\frac{d-N+1}{2}+s)\Gamma(s+1) }
{ 2\Gamma(\frac{d-N+1}{2}) } 
\left(\frac{1}{R_N }\right)^s 
\times\\
&&\hspace{1.0cm}\times     
\sum\limits_{k=1}^N 
\left(\frac{\partial_k R_N }{R_N} 
\right) \;
{\bf k}^- 
\; J_{N}(d+2s; \{p_ip_j\}, \{m_i^2\}), 
\nonumber
\end{eqnarray}
for $i,j=1,2,\cdots, N$ and 
$\partial_k=\partial/\partial m_k^2$.
Here the operator 
${\bf k}^-$ is defined as~\cite{Phan:2018cnz}
\begin{eqnarray}
{\bf k}^-  J_{N}(d; \{p_ip_j\}, \{m_i^2\}) 
&=& \int \frac{d^d k}{i \pi^{d/2}} \frac{1}
{P_1 P_2\dots P_{k-1}P_{k+1} 
\dots P_{N-1}P_N}.
\end{eqnarray}
The relation (\ref{JNJN1}) 
indicates that the integral $J_N$ can be constructed
by taking one-fold Mellin-Barnes (MB) 
integration over 
$J_{N-1}$ in $d+2s$. This representation 
has several advantages. First, analytic formulas 
for $J_N$ can be derived from basic functions 
which are scalar one-loop one-point functions. 
Second, $J_N$ is expressed as functions of 
kinematic variables such as $m_j^2$ for 
$j=1,2,\cdots, N$ 
and $R_N$. As a consequence of this fact, 
analytic expressions for $J_N$
reflect the symmetry 
as well as threshold behavior of the 
corresponding Feynman topologies. 
Two special cases of (\ref{JNJN1})
are also mentioned as follows:  
\begin{enumerate}
\item $Y_N \rightarrow 0 $ 
and $G_{N-1} \neq 0$:
In this case, $R_N \rightarrow 0$ and 
we have~\cite{Fleischer:2003rm}
(deriving this equation for $N=4$ is 
shown in the appendix $D$)
\begin{eqnarray}
\label{mijkl0-sol}
J_N(d;\{p_ip_j\}, \{m_i^2\}) =
\frac{1}{d-N-1} \sum\limits_{k=1}^N
\left(\frac{\partial_k Y_N}{G_{N-1}} \right)
{\bf k^{-} } J_N(d-2;\{p_ip_j\}, \{m_i^2\}).
\end{eqnarray}
\item $G_{N-1} \rightarrow 0$ and $Y_N \neq 0$:  %
In this case, $R_N \rightarrow \infty$. 
We close the integration contour
in (\ref{JNJN1}) to the right. 
Taking residue contributions
from poles of $\Gamma(\cdots -s)$. 
In the limit $R_N \rightarrow \infty$,
we find only the term with $s=0$ is 
non-zero. The result then reads
\begin{eqnarray}
\label{GijkN0}
J_N(d;\{p_ip_j\}, \{m_i^2\}) 
= -\frac{1}{2} \sum\limits_{k=1}^N   
\left( \frac{\partial_k Y_N }{Y_N}  \right) \;
{\bf k}^{-}  J_N(d;\{p_ip_j\}, \{m_i^2\}).
\end{eqnarray}
This equation is equivalent to ($65$) 
in \cite{Devaraj:1997es} and ($3$) 
in \cite{Fleischer:2003rm}. 
\end{enumerate}

We turn our attention to apply the method 
for evaluating scalar one-loop Feynman 
integrals. The detailed evaluations 
for scalar one-loop two-, three- and four-point 
functions are presented in next 
sections. As we pointed out in this
section, the prescription 
$i\rho$ always follows with 
$R_N$ as $R_N-i\rho$. In order to 
simplify the notation,
we omit $i\rho$ in $R_N$ in the next 
calculations. This term puts back
into the final results 
when it is necessary. 
\section{One-loop two-point functions}
The master equation for $J_2$ is 
obtained by setting $N=2$ in~(\ref{JNJN1}). 
MB representation for $J_2$ then reads
\begin{eqnarray}
\label{MBJ2}
J_2 \equiv J_2(d; p^2, m_1^2, m_2^2)  &=& 
\frac{1}{2\pi i} 
\int\limits_{-i\infty}^{+i\infty}ds \; 
\dfrac{\Gamma(-s)
\Gamma\left( \frac{2-d}{2}-s\right)
\Gamma\left(\frac{d-1}{2} +s\right)
\Gamma(s+1) }
{2\Gamma\left(\frac{d-1}{2} \right) }
\times \\
&&\hspace{1cm}
\times
\left(\dfrac{1}{R_2 } \right)^{s} 
\left\{ 
\left(\dfrac{\partial_2 R_2 }{R_2}\right)  
(m_1^2)^{\frac{d-2}{2} +s} 
+ (1 \leftrightarrow 2)
\right\}. 
\nonumber
\end{eqnarray}
Note that we used the analytic formula for $J_1$ 
in~\cite{tHooft:1978jhc} with $d$ shifted
to $d\rightarrow d+2s$. We write $J_1$
in $d+2s$ explicitly as follows:
\begin{eqnarray}
J_1(d+2s; m^2) &=&-
\Gamma \left(\frac{2-d}{2}-s\right) 
(m)^{\frac{d-2}{2}+s}.
\end{eqnarray}
In order to evaluate 
the MB integrals in (\ref{MBJ2}), 
we close the integration contour
to the right. The residue contributions
to $J_2$ at the sequence 
poles of $\Gamma(-s)$ and 
$\Gamma\left(\frac{2-d}{2}-s\right)$
are taken into account. 

First, we calculate the residue 
at the poles of $\Gamma(-s)$. 
In this case, 
$s = m$ for $m=0,1, \cdots, \mathbb{N}$. 
Subsequently, we can apply the 
reflect formula for gamma functions
(\ref{reflect}) in the appendix $B$. 
In detail, it is implied that 
\begin{eqnarray}
\label{reflect1}
\Gamma\left( \frac{2-d}{2}-s\right)
\Gamma\left(\frac{d}{2}+s\right)
=-(-1)^s \Gamma\left(\frac{4-d}{2}\right) 
\Gamma\left(\frac{d-2}{2}\right).
\end{eqnarray}
With the help of (\ref{reflect1}), 
the MB representation 
in (\ref{MBJ2}) is casted into 
the form of
\begin{eqnarray}
\label{j2sm-a}
\dfrac{J_2}
{\Gamma\left(\frac{4-d}{2} \right)}
\Big|_{s=m}
&=& 
-\dfrac{\Gamma\left(\frac{d-2}{2} \right) }
{2\;\Gamma\left(\frac{d-1}{2} \right) }\; 
\frac{1}{2\pi i}
\int\limits_{-i\infty}^{+i\infty}ds \; 
\dfrac{\Gamma(-s)
\;\Gamma\left(\frac{d-1}{2} +s\right)
\;\Gamma(s+1) }
{\Gamma\left(\frac{d}{2}+s \right) }\; 
\times \\
&&\hspace{2cm} 
\times \left\{ 
\left(\frac{\partial_2 R_2 }{R_2}\right)  
(m_1^2)^{\frac{d-2}{2}} 
\left(-\dfrac{m_1^2}{R_2} \right)^{s}  
+ (1 \leftrightarrow 2)
\right\}. 
\nonumber
\end{eqnarray}
Using ($1.6.1.6$) in Ref.~\cite{Slater}, 
these MB integrals are 
expressed
in terms of Gauss hypergeometric 
series as follows:
\begin{eqnarray}
\label{jsm-b}
\dfrac{J_2}
{\Gamma\left( \frac{4-d}{2} \right)} 
\Big|_{s=m}
&=& 
-\dfrac{\Gamma\left(\frac{d-2}{2} \right) }
{2\;\Gamma\left(\frac{d}{2} \right) }\; 
\left\{ 
\left(\frac{\partial_2 R_2 }{R_2} \right)
(m_1^2)^{\frac{d-2}{2}}\;
\Fh21\Fz{1, \frac{d-1}{2} }{ \frac{d}{2} }
{\frac{m_1^2}{R_2}} 
 + (1 \leftrightarrow 2)
\right\},  
\end{eqnarray}
provided that  $\left|m_1^2/R_2\right|<1$, 
$\left|m_2^2/R_2\right|<1$ and 
$\mathcal{R}$e$(d-2)>0$. Using ($1.3.15$)
in \cite{Slater}, we arrive at another 
representation for 
(\ref{jsm-b}) 
\begin{eqnarray}
\label{J2sm-c}
\dfrac{J_2}
{\Gamma\left( \frac{4-d}{2} \right)} 
\Big|_{s=m}
&=& - \dfrac{\Gamma( \frac{d-2}{2} )}
{2 \Gamma(\frac{d}{2}) }
\left \{ 
\left(
\frac{\partial_2 R_2 }{R_2} \right)
\frac{ (m_1^2)^{\frac{d-2}{2}} }
{\sqrt{1-m_1^2/R_2 } }
\;\Fh21\Fz{\frac{d-2}{2}, \frac{1}{2} } 
{\frac{d}{2}}{ \dfrac{m_1^2}{R_2} }  
+  (1 \leftrightarrow 2)    
\right \},  
\end{eqnarray}
provided that  $\left|m_1^2/R_2\right|<1$, 
$\left|m_2^2/R_2\right|<1$ and 
$\mathcal{R}$e$(d-2)>0$.

We next consider 
the residue at the second sequence poles of 
$\Gamma\left(\frac{2-d}{2} -s\right)$. 
In this case, $s =\frac{2-d}{2}+m $ for 
$m \in \mathbb{N}$. These contributions 
read
\begin{eqnarray}
\label{J2minus-sm}
J_2\Big|_{s =\frac{2-d}{2}+m}  
&=& \sum_{m=0}^{\infty} \dfrac{(-1)^m}{m!}
\dfrac{\Gamma(\frac{d-2}{2}-m ) 
\Gamma( \frac{4-d}{2} +m ) 
\Gamma(m+\frac{1}{2})}
{2\Gamma(\frac{d-1}{2}) } \\
&&\hspace{0.7cm}
\times 
(R_2)^{\frac{d-2}{2} }\;
\left[
\left(\frac{\partial_2 R_2 }{R_2} \right) 
\left( \dfrac{m_1^2}{R_2}\right)^m 
+ \left(
\frac{\partial_1 R_2 }{R_2} 
\right) 
\left( \dfrac{m_2^2}{R_2}\right)^m
\right] \nonumber\\
&&\hspace{-2cm}=  
\dfrac{\Gamma(\frac{d-2}{2} )
\Gamma( \frac{4-d}{2} ) }
{2\Gamma(\frac{d-1}{2}) } 
(R_2)^{\frac{d-2}{2} } 
\sum_{m=0}^{\infty} 
\dfrac{\Gamma(m+\frac{1}{2})}
{\Gamma(m+1)}
\left[
\left(
\frac{\partial_2 R_2 }{R_2} 
\right) 
\left( \dfrac{m_1^2}{R_2}\right)^m 
+ 
\left(
\frac{\partial_1 R_2 }{R_2} 
\right) 
\left( \dfrac{m_2^2}{R_2}\right)^m
\right]
\label{J2minus-s-b}\\
&&\hspace{-2cm}= 
\dfrac{\sqrt{\pi}}{2}\;
\dfrac{\Gamma(\frac{4-d}{2} ) 
\Gamma(\frac{d-2}{2}) }
{2\Gamma(\frac{d-1}{2})} 
\left(R_2\right) ^{\frac{d-4}{2}} 
\left[ 
\dfrac{\partial_2 R_2}{\sqrt{1-m_1^2/R_2 }} 
+ 
\dfrac{\partial_1 R_2}{\sqrt{1-m_2^2/R_2 }}  
\right]. 
\label{J2minus-s-c}
\end{eqnarray}
Noting that from (\ref{J2minus-sm}) 
to (\ref{J2minus-s-b}), 
we have already applied the reflect 
formulas for gamma functions 
(see (\ref{reflect}) in appendix $A$ 
for more detail). 
Summing all the above contributions 
in Eqs.~(\ref{J2sm-c},~\ref{J2minus-s-c}),
we finally get 
\begin{eqnarray}
\label{J2F21-final}
\dfrac{J_2}{\Gamma(\frac{4-d}{2} ) }   
&=& 
\dfrac{\sqrt{\pi}}{2} 
\dfrac{ \Gamma(\frac{d-2}{2}) }
{\Gamma(\frac{d-1}{2} ) }
\left(R_2\right) ^{\frac{d-4}{2}} 
\left[  
\dfrac{ \partial_2 R_2}
{\sqrt{1-m_1^2/R_2}}  
+
\dfrac{ \partial_1 R_2}
{\sqrt{1-m_2^2/R_2}} 
\right]
\\
&& 
-\dfrac{ \Gamma( \frac{d-2}{2} )} 
{2\Gamma(\frac{d}{2}) }
\left\{ 
\left(\frac{\partial_2 R_2 }{R_2} \right) 
\dfrac{(m_1^2)^{\frac{d-2}{2}} }
{\sqrt{1- m_1^2/R_2 }}
\Fh21\Fz{\frac{d-2}{2}, \frac{1}{2} } 
{\frac{d}{2}}{ \dfrac{m_1^2}{R_2} }  
+ 
(1\leftrightarrow2)
\right\}, \nonumber
\end{eqnarray}
provided that $\left|m_1^2/R_2\right|<1$, 
$\left|m_2^2/R_2\right|<1$ and 
$\mathcal{R}$e$(d-2)>0$. 
Eq.~(\ref{J2F21-final})
is unchanged with exchanging 
$m_1^2 \leftrightarrow m_2^2$. This 
reflects the symmetry of the scalar 
one-loop two-point Feynman diagrams. 
The result in (\ref{J2F21-final}) 
has shown in~\cite{Bluemlein:2017rbi}
and gives fully agreement with
\cite{Fleischer:2003rm}. 
It is an important to remark that the 
solution for $J_2$ in 
(\ref{J2F21-final}) is also valid when 
$d\rightarrow d+2n$ for $n\in \mathbb{N}$.
We would like to stress that 
one can perform the analytic continuation 
for $J_2$ in (\ref{J2F21-final}) 
to extend the kinematic regions 
for one-loop two-point functions. 
The analytic continuation 
formulas for Gauss hypergeometric functions 
are given from (\ref{f21c}) to 
(\ref{z-}) in the appendix $B$. 
As an example,
using (\ref{f21f}), the result reads
\begin{eqnarray}
\label{b7b}
\dfrac{J_2}{\Gamma\left(\frac{4-d}{2} \right) } 
&=&\left( \frac{\partial_2 R_2}{R_2} 
\right)
(m_1^2)^{\frac{d-2}{2}}
\Fh21\Fz{ 1, \frac{d-1}{2} } 
{\frac{3}{2}}{1-\frac{m_1^2}{R_2} }
+
(1\leftrightarrow2). 
\end{eqnarray}
With (\ref{f21b}), we arrive at
\begin{eqnarray}
\label{b7a}
\dfrac{J_2}{\Gamma\left(\frac{4-d}{2} \right) } 
&=&\left( \frac{\partial_2 R_2}{R_2} 
\right)
R_2^{\frac{d-2}{2}}
\Fh21\Fz{\frac{4-d}{2}, \frac{1}{2} } 
{\frac{3}{2}}{1-\frac{m_1^2}{R_2} }
+
(1\leftrightarrow2). 
\end{eqnarray}

We are going to consider special 
cases for scalar one-loop two-point 
Feynman integrals. 
\subsection{$G_1 \neq 0$        
and $R_2 = 0$}                    
If $R_2 =0$, we verify that  
$p^2 = (m_1-m_2)^2$, or $(m_1+m_2)^2$. 
Applying (\ref{mijkl0-sol}) for 
$N=2$ the result reads
\begin{eqnarray}
\label{j2m120}
\dfrac{J_2}{\Gamma(\frac{4-d}{2}) }  
&=& 
\frac{1}{d-3} \left[      
\frac{(m_1^2)^{\frac{d-3}{2}}
}{(m_1\pm m_2)^3}
+ (1 \leftrightarrow 2)       
\right].
\end{eqnarray}
In the limit of $m_1\rightarrow m_2 =m$ and 
$p^2 =4m^2$,  the result reads
\begin{eqnarray}
J_2 &=& 
\frac{\Gamma(\frac{4-d}{2})}{2(d-3)}
(m^2)^{\frac{d-6}{2}}.
\end{eqnarray}
\subsection{$G_1 = 0$} 
Following (\ref{GijkN0}) for 
$N=2$, we arrive at
\begin{eqnarray}
\label{J2G20}
J_2 &=& \Gamma\left(\frac{4-d}{2} \right)
(m_2^2)^{\frac{d-4}{2}}\;
\Fh21\Fz{\frac{4-d}{2}, 1}{2}
{1-\frac{m_1^2}{m_2^2}} 
= \Gamma\left( \frac{2-d}{2} \right)
\dfrac{ (m_2^2)^{\frac{d-2}{2}}-
(m_1^2)^{\frac{d-2}{2}} }{m_1^2-m_2^2}.   
\end{eqnarray}
If $m_1^2 =m_2^2$, one presents $J_2$ as
\begin{eqnarray}
J_2 &=& \Gamma\left(\frac{4-d}{2} \right) 
(m^2)^{\frac{d-4}{2}}. 
\end{eqnarray}
\subsection{$R_2= m_1^2$ %
or $R_2=m_2^2$} 
For the case of $R_2= m_1^2$ or 
$R_2=m_2^2$, one relies on $(\ref{b7a})$. 
If $R_2 =m_1^2= m_2^2=m^2$, 
the result in (\ref{b7a}) 
simplifies to
\begin{eqnarray}
J_2  = \Gamma\left(\frac{4-d}{2} \right) 
(m^2)^{\frac{d-4}{2}}. 
\end{eqnarray}
Under the condition 
$\mathcal{R}$e$(d-4)>0$,
when $m^2 \rightarrow 0$, one then 
gets $J_2=0$. 
\subsection{$m_1^2=m_2^2=0$}
If one of $m_i^2=0$, for $i=1,2$, 
we rely on (\ref{J2F21-final}). 
In the case of $m_1^2=m_2^2=0$,  
from (\ref{J2F21-final}) we get
\begin{eqnarray}
\label{2point-fey-mij0}
J_2&=& \frac{\sqrt{\pi}}{2}
\frac{ \Gamma\Big(\frac{4-d}{2}\Big)
\Gamma\Big(\frac{d-2}{2}\Big)}
{\Gamma\left(\frac{d-1}{2}\right)} 
\Big(-\frac{p^2}{4} \Big)^{ \frac{d-4}{2}},
\end{eqnarray}
provided that 
$\mathcal{R}$e$\left(d-2\right)>0$. 
We note that
$p^2$ means $p^2 + i\rho$. 
Therefore, if $p^2>0$ the term 
$\left(-p^2/4\right)^{\frac{d-4}{2}}$ 
is well-defined.
\section{One-loop three-point functions}
Setting $N=3$ in (\ref{JNJN1}), 
master equation for $J_3$ reads
\begin{eqnarray}
\label{J3J31}
J_3 \equiv
J_3(d; \{p_i^2\}, \{m_i^2\}) 
&=& -\dfrac{1}{2\pi i} 
\int\limits_{-i\infty}^{+i\infty}ds \; 
\dfrac{\Gamma(-s)\; 
\Gamma(\frac{d-2}{2}+s)\Gamma(s+1) }
{ 2\Gamma(\frac{d-2}{2}) } 
\left(\frac{1}{R_3}\right)^s \times 
\\
&&\hspace{1cm}\times     \sum\limits_{k=1}^3 
\left( 
\frac{\partial_k R_3 }{R_3} \right) \;
{\bf k}^- J_3(d+2s; \{p_i^2\}, \{m_i^2\}), 
\nonumber
\end{eqnarray}
for $i=1,2,3$. The term 
${\bf k}^- J_3(d+2s; \{p_ip_j\}, \{m_i^2\})$ 
becomes scalar one-loop two-point 
functions by shrinking an propagator 
$k$-th in the integrand of $J_3$. 
In the next steps, we take 
the contour integrals in (\ref{J3J31}). 
In order to understand how to take the 
contour integrals, we chose the term 
with $k=3$ in (\ref{J3J31}) for 
illustrating. This term is written 
explicitly as follows:  
\begin{eqnarray}
\label{L123}
J_{3,(123)} &=& -\frac{1}{2\pi i} 
\int\limits_{-i\infty}^{+i\infty}
ds \dfrac{ \Gamma(-s)\;\Gamma(s+1) 
\Gamma\left(\frac{d-2}{2}+s\right) }
{ 2\;\Gamma(\frac{d-2}{2}) }
\left(\frac{1}{R_3}\right)^{s}
\left(\frac{\partial_3 R_3}{R_3}\right) 
J_2(d+2s; p_1^2, m_1^2, m_2^2)
\nonumber\\
&&\\
\label{J3at123}
&=& -\frac{1}{2\pi i} 
\int\limits_{-i\infty}^{+i\infty}
ds \dfrac{\sqrt{\pi}\Gamma(-s)
\Gamma(s+1) 
\Gamma\left(\frac{d-2}{2}+s\right)
\Gamma(\frac{4-d}{2} -s )
\Gamma(\frac{d-2}{2} +s )  }
{4\Gamma(\frac{d-2}{2}) 
\Gamma(\frac{d-1}{2} +s) }
\times \nonumber\\
&& \hspace{1.cm}
\times 
\left(\frac{\partial_3 R_3}{R_3}\right) 
\left[  
\dfrac{ \partial_2 R_{12} }
{\sqrt{1-m_1^2/R_{12} }}  
+
\dfrac{ \partial_1 R_{12} }
{\sqrt{1-m_2^2/R_{12} }} 
\right]
\;
\left(R_{12}\right) ^{\frac{d-4}{2}} 
\left( -\frac{R_{12}}{R_3}\right)^s
\\
&&
+\frac{1}{2\pi i} 
\int\limits_{-i\infty}^{+i\infty}
ds \dfrac{\Gamma(-s)\;\Gamma(s+1) 
\Gamma\left(\frac{d-2}{2}+s\right)
\Gamma(\frac{4-d}{2} -s )
\Gamma( \frac{d-2}{2}+s )
}
{4\;\Gamma(\frac{d-2}{2})
\Gamma(\frac{d}{2} +s) 
}
\left(\frac{\partial_3 R_3}{R_3}\right) 
\times
\nonumber\\
&& \hspace{1cm}
\times 
\left\{ 
\left(
\frac{\partial_2 R_{12} }{R_{12} } 
\right) 
\dfrac{(m_1^2)^{\frac{d-2}{2}} }
{\sqrt{1- m_1^2/R_{12} }} 
\Fh21\Fz{\frac{d-2}{2}+s, \frac{1}{2} } 
{\frac{d}{2}+s }{ \dfrac{m_1^2}{R_{12} } } 
\left(- \dfrac{m_1^2}{R_3} \right)^s
+  (1\leftrightarrow2)
\right\}. \nonumber
\end{eqnarray}
For taking the MB 
integrals in (\ref{J3at123}), one closes 
the integration contour to the right. 
The residue contributions at the poles 
of the $\Gamma(\cdots -s)$ are taken 
into account.

First, the contributions of residua at 
the poles $s = m$ with $m\in \mathbb{N}$. 
In this case, one first applies the 
reflect formula~(\ref{reflect}) for 
gamma function:
\begin{eqnarray}
\Gamma\left( \frac{4-d}{2} -s \right)\; 
\Gamma \left(\frac{d-2}{2} +s \right) = 
(-1)^s \; \Gamma\left( \frac{4-d}{2} \right)
\;\Gamma \left(\frac{d-2}{2}\right).
\end{eqnarray}
Using this identity, the first MB
integration reads
\begin{eqnarray}
\dfrac{J_{3,(123)}^{\mathrm{1-term}} }
{\Gamma\left(\frac{4-d}{2} \right) }
\Big|_{s=m}
&=& -\frac{\sqrt{\pi}}{4}
\left(
\frac{\partial_3 R_3}{R_3} 
\right) 
\left[ \frac{\partial_2 R_{12} }
{\sqrt{1- m_1^2/R_{12} }}
+ \frac{\partial_1 R_{12} }
{\sqrt{1- m_2^2/R_{12} }}  \right]
\left(R_{12}\right)^{\frac{d-4}{2}}  
 \times  \\
&&\hspace{1.7cm} \times
\frac{1}{2\pi i} 
\int\limits_{-i\infty}^{+i\infty}\;ds \;
\frac{ \Gamma(-s)\;\Gamma(s+1) 
\Gamma\Big(\frac{d-2}{2}+s\Big) }
{ \Gamma(\frac{d-1}{2}+s)}
\left(-\frac{R_{12}}
{R_3 }  \right)^s. 
\nonumber
\end{eqnarray}
This MB integral
is then expressed
in terms of hypergeometric $_2F_1$ as
follows:
\begin{eqnarray}
\label{j3d1}
\dfrac{J_{3,(123)}^{\mathrm{1-term}} }
{\Gamma\left(\frac{4-d}{2} \right) }
\Big|_{s=m}
&=& -\frac{\sqrt{\pi}}{4}
\frac{\Gamma \left(\frac{d-2}{2}\right) }
{\Gamma \left( \frac{d-1}{2} \right)}
\left(
\frac{\partial_3 R_3}{R_3} 
\right)  
\left[ 
\frac{\partial_2 R_{12} }
{\sqrt{1-m_1^2/R_{12} }}
+ 
(1\leftrightarrow2)
\right]
\left(R_{12}\right)^{\frac{d-4}{2}} 
\Fh21\Fz{ \frac{d-2}{2}, 1}
{\frac{d-1}{2} }
{\frac{R_{12} }{R_3 } }, 
\nonumber\\
\end{eqnarray}
provided that $\left|R_{12}/R_3\right|<1$ 
and $\mathcal{R}$e$\left(d-3\right)>0$.
The second MB integral reads
\begin{eqnarray}
\hspace{0cm}
\dfrac{J_{3,(123)}^{\mathrm{2-term}}}
{\Gamma\left( \frac{4-d}{2}\right)} 
\Big|_{s=m}
&=&  
\frac{1}{2\pi i} 
\int\limits_{-i\infty}^{+i\infty}\;ds \;
\dfrac{\Gamma(-s)\;
\Gamma(s+1) \Gamma\Big(\frac{d-2}{2}+s\Big) }
{\Gamma(\frac{d}{2}+s)} 
\left(\frac{\partial_3 R_3}{4R_3} \right)
\times \\
&&\hspace{-0.5cm} \times \left\{ 
\left(\dfrac{\partial_2 R_{12} }
{R_{12} }\right)
\dfrac{(m_1^2)^{\frac{d-2}{2} } }
{ \sqrt{1-m_1^2/R_{12} }}
\left(-\frac{m_1^2}{R_3} \right)^s \;
\Fh21\Fz{\frac{d-2}{2} +s, \frac{1}{2} } 
{\frac{d}{2} +s}{ \dfrac{m_1^2}
{R_{12}} } +  (1\leftrightarrow 2)    
\right \} 
\nonumber\\
&&\hspace{-2.8cm}
=
\label{j3d2}
\dfrac{\Gamma\left(\frac{d-2}{2} \right)  } 
{4\Gamma\left(\frac{d}{2}\right) }
\left(
\frac{\partial_3 R_3}{R_3} 
\right) 
\left[ 
\left(\dfrac{\partial_2 R_{12} }
{R_{12} } \right)
\dfrac{(m_1^2)^{\frac{d-2}{2} } }
{\sqrt{1- m_1^2/R_{12} }} 
F_1 \left(\dfrac{d-2}{2}; 1,
\frac{1}{2}; \frac{d}{2} ; 
\frac{m_1^2}{R_3}, 
\dfrac{m_1^2}{R_{12} } \right)
+  (1\leftrightarrow 2)
\right],               
\nonumber\\
\end{eqnarray}
provided that $\left|m_{(1,2)}^2/R_3\right|$, 
$\left|m_{(1,2)}^2/R_{12} \right|<1$ and 
$\mathcal{R}$e$\left(d-2\right)>0$.

In the next steps, 
the residue contributions 
at the second sequence poles
$s = \frac{4-d}{2}+ m$ for 
$m\in \mathbb{N}$ are
taken into account. The next MB 
integrations are considered 
as follows:
\begin{eqnarray}
J_{3,(123)}^{\mathrm{1-term}}\Big|_
{s=\frac{4-d+2m}{2} } &=&
- \frac{\sqrt{\pi}  }{2\pi i} 
\int\limits_{-i\infty}^{+i\infty}\;ds \;
\dfrac{\Gamma(-s)\;
\Gamma(s+1) \Gamma^2\Big(\frac{d-2}{2}+s\Big) 
\Gamma\left(\frac{4-d}{2} -s\right) 
}{4\;\Gamma\left( \frac{d-2}{2}\right)
\Gamma(\frac{d-1}{2}+s)}  
\\
&& \hspace{0cm}
\times 
\left( 
\frac{\partial_3 R_3}{R_3} 
\right)
\left[\dfrac{\partial_2 R_{12} }
{\sqrt{1 - m_1^2/R_{12} }}
+ \dfrac{\partial_1 R_{12} }
{\sqrt{1-m_2^2/R_{12} }} \right]
\left(R_{12} \right)^{\frac{d-4}{2}}  
\left(-\frac{R_{12} }{R_3} \right)^s
\nonumber
\\
&=&\left(\frac{\partial_3 R_3}{2\; R_3} \right)  
\left[ \dfrac{\partial_2 R_{12} }
{\sqrt{1-m_1^2/R_{12} }}
 + 
\dfrac{\partial_1 R_{12} }
{\sqrt{1-m_2^2/R_{12} }}
\right] (R_3 )^{\frac{d-4}{2}} 
\; \Fh21\Fz{1, 1}{\frac{3}{2}}
{\frac{R_{12} }{R_3 }},
\nonumber\\
\label{j3bound1}
&& \\
\dfrac{J_{3,(123)}^{\mathrm{2-term}}}
{\Gamma\left( \frac{4-d}{2}\right)} 
\Big|_{s=\frac{4-d}{2} + m}
&=& -
\left(
\frac{\partial_3 R_3}{2R_3} \right) 
\left[ 
\left(\frac{\partial_2R_{12} }
{2R_{12} }\right)
\frac{m_1^2(R_3)^{\frac{d-4}{2}}}
{\sqrt{1- m_1^2/R_{12} }} 
F_1 \left( 1; 1, \frac{1}{2}; 2;
\frac{m_1^2}{R_3}, 
\dfrac{m_1^2}{R_{12} } \right) 
+  (1\leftrightarrow 2)
\right]
, \label{j3bound2}
\nonumber\\
\end{eqnarray}
provided that $\left|m_{(1,2)}^2/R_3\right|$, 
$\left|m_{(1,2)}^2/R_{12}\right|<1$
and $\left|R_{12}/R_3\right|<1$.

Summing all the above contributions, the final 
result for $J_3$ is written as a compact form
\begin{eqnarray}
\label{J3normal}
\dfrac{J_3}
{\Gamma\left( \frac{4-d}{2}\right) }   
&=&
-J_{123}^{(d=4)}\;
(R_3 )^{\frac{d-4}{2}} + J_{123}^{(d)} 
\nonumber\\
&& 
+ \Big\{(1,2,3) \leftrightarrow (2,3,1)\Big\} 
\\
&&
+  \Big\{(1,2,3) 
\leftrightarrow (3,1,2)\Big\}. 
\nonumber
\end{eqnarray}
Where
$J_{123}^{(d)}$ is obtained 
from (\ref{j3d1}) and (\ref{j3d2}).
It is given by
\begin{eqnarray}
\label{J123normal}
J_{123}^{(d)} &=& - \dfrac{\sqrt{\pi} 
\Gamma\left(\frac{d-2}{2}\right)  }
{4\Gamma\left(\frac{d-1}{2}\right) }
\left(\frac{\partial_3 R_3}{R_3}\right)  
\Bigg[
\dfrac{\partial_2 R_{12} }
{\sqrt{1-m_1^2/R_{12} }} +  
(1\leftrightarrow 2) 
\Bigg]
\left(R_{12}\right)^{\frac{d-4}{2}} 
\Fh21\Fz{\frac{d-2}{2}, 1}
{\frac{d-1}{2} }
{ \dfrac{R_{12} }{R_3 } }                     
\\ 
&&
+
\dfrac{\Gamma\left(\frac{d-2}{2}\right)}
{4\Gamma\left(\frac{d}{2}\right) }
\left(\frac{\partial_3 R_3}{R_3} \right)  
\Bigg[
\left( \dfrac{\partial_2 
R_{12} }{R_{12} } \right)
\dfrac{(m_1^2)^{\frac{d-2}{2} } }
{\sqrt{1-m_1^2/R_{12}}} 
F_1 \left(\dfrac{d-2}{2}; 
1, \frac{1}{2}; \frac{d}{2}; 
\frac{m_1^2}{R_3}, 
\dfrac{m_1^2}{R_{12} } \right)  
+ (1\leftrightarrow 2)      
\Bigg],  
\nonumber
\end{eqnarray}
provided that 
$\left|m_{i}^2/R_{ij} \right|<1$, 
$\left|R_{ij}/R_{ijk} \right|<1$
for $i,j,k=1,2,3$ and 
$\mathcal{R}$e$\left(d-2 \right)>0$. 
The latter condition 
always meets when {$d>2$}. 
The kinematic variables 
$R_{ijk}, R_{ij}$ and 
$m_{i}$ for $i,j,k =1,2,3$, etc., 
may not satisfy the former conditions. 
If the absolute value 
of the arguments of {$\Fh21$} 
and {the Appell functions $F_1$} 
in (\ref{J123normal}) 
{are larger than one,} 
we have to perform analytic continuations 
for these functions as in~\cite{Slater,olsson}.
The result for $J_3$ has been shown in
\cite{Bluemlein:2017rbi,Phan:2018cnz}.
The term $J_{123}^{(d=4)}$ is obtained
from (\ref{j3bound1}, \ref{j3bound2})
or taking $d\rightarrow 4$ of 
(\ref{J123normal}). This term is given 
\begin{eqnarray}
\label{J123normal4}
 J_{123}^{(d=4)} &=& - 
 \left(\frac{\partial_3 R_3}{2R_3} \right)  
\Big[
\dfrac{\partial_2 R_{12} }
{\sqrt{1-m_1^2/R_{12} }} +  
(1\leftrightarrow 2) 
\Big]
\Fh21\Fz{ 1, 1}{3/2}
{ \dfrac{R_{12} }{R_3 } }                     
\\ 
&& +
\left(\frac{\partial_3 R_3}{2R_3} \right)  
\left[
\left( \dfrac{\partial_2 
R_{12} }{2R_{12} } \right)
\dfrac{m_1^2}
{ \sqrt{1-m_1^2/R_{12} } } \; 
F_1 \left(1; 1, \frac{1}{2}; 2; 
\frac{m_1^2}{R_3}, 
\dfrac{m_1^2}{R_{12} } \right)  
+  (1\leftrightarrow 2)     
\right].
\nonumber
\end{eqnarray}

We emphasis that the solution 
$(\ref{J3normal})$ 
for $J_3$ with 
$(\ref{J123normal})$ is equivalent
to ($74$) in Ref.~\cite{Fleischer:2003rm}. 
But the terms $J^{(d=4)}_{123}, \cdots$ in our solution 
cover the condition ($73$) in 
Ref.~\cite{Fleischer:2003rm}.
Since the boundary term given 
in ($74$) of Ref.~\cite{Fleischer:2003rm} 
was obtained by
asymptotic theory of complex Laplace-type
integrals. 
This term is only valid
in a kinematic sub-domain in which the 
asymptotic theory of Laplace-type
can be applied.
The analytic continuation
for the boundary term 
in~\cite{Fleischer:2003rm}
has not been discussed. 
We provide a 
complete analytic solution for $J_3$ in 
comparison with ($74$) in 
Ref.~\cite{Fleischer:2003rm}. We refer 
to our 
previous work \cite{Bluemlein:2017rbi} 
in which the numerical studies for 
this problem have discussed.

One-fold integral 
and all transformations for $F_1$ can be 
found in appendix $B$. 
Applying the relation~(\ref{f1relation})
for $F_1$ in appendix $C$, we arrive at
another representation for 
(\ref{J123normal}, \ref{J123normal4}):
\begin{eqnarray}
J_{123}^{(d)} &=&
\dfrac{
\left(\partial_3 R_3\right)       
\left(\partial_2 R_{12} \right) 
}{2(m_1^2-R_3) }  
(m_1^2)^{\frac{d-4}{2} } 
F_1 \left(1; \frac{4-d}{2}, 1; \frac{3}{2};
1-\frac{R_{12} }{m_1^2}, 
\dfrac{R_{12}-m_1^2}
{R_3 - m_1^2}  \right)  
+  (1\leftrightarrow 2), 
\label{J123rotaion}
\\
J_{123}^{(d=4)} &=&
\dfrac{
\left(\partial_3 R_3\right)       
\left(\partial_2 R_{12} \right) 
}{2(m_1^2-R_3) } 
\Fh21\Fz{ 1, 1}{\frac{3}{2}}{ 
\dfrac{R_{12}-m_1^2}
{R_3 - m_1^2} 
}
+  (1\leftrightarrow 2),
\label{J123rotaion-boundary}
\end{eqnarray}
provided that the absolute value of 
the arguments of {$\Fh21$}
and {the Appell functions $F_1$} in 
this presentation are less 
than $1$. 
\subsection{Massless internal lines} 
For the massless case, under the condition 
$\mathcal{R}$e$(d-2)>0$,
all terms related to 
Appell $F_1$ functions in (\ref{J3normal}) 
vanish, the result 
then reads
\begin{eqnarray}
\dfrac{J_3}
{\Gamma\left( \frac{4-d}{2}\right) } 
&=& 
-  \; (R_3)^{\frac{d-4}{2}}\; 
\left( 
\frac{\partial_3 R_3}{2R_3} 
\right)
\Big|_{m_i^2\rightarrow0}
\Fh21\Fz{1, 1}{\frac{3}{2}}
{ \dfrac{R_{12} }{R_3 } }     \\
&&+ \dfrac{\sqrt{\pi}
\Gamma\left(\frac{d-2}{2}\right) } 
{4\;\Gamma\left(\frac{d-1}{2}\right) }
\left(\frac{\partial_3 R_3}{R_3} \right)
\Big|_{m_i^2\rightarrow0}\;
(R_{12})^{\frac{d-4}{2}}\;
\Fh21\Fz{ \frac{d-2}{2}, 1}
{\frac{d-1}{2} }
{ \dfrac{R_{12} }{R_3 } }         
\nonumber\\
&& \nonumber\\
&& 
+ \Big\{ (1,2,3) \rightarrow (2,3,1) \Big\} \quad 
+ \quad \Big\{ (1,2,3) \rightarrow (3,1,2) \Big\}. 
\nonumber
\end{eqnarray}
In order to cross check with the result 
in~\cite{Davydychev:1999mq}, we write $J_3$ as a 
function of $p_1^2, p_2^2, p_3^2$ explicitly
\begin{eqnarray}
\dfrac{J_3}{\Gamma\left( \frac{4-d}{2}\right) }
&=& -\left( \dfrac{p_2^2 +p_3^2 -p_1^2}
{2\;p_2^2p_3^2} \right)\;
\left(
\dfrac{p_1^2p_2^2p_3^2}
{\lambda(p_1^2, p_2^2, p_3^2)}  
\right)^{ \frac{d-4}{2}}
\Fh21\Fz{1,1}{\frac{3}{2}} 
{- \dfrac{\lambda(p_1^2, p_2^2, p_3^2)}
{4p_2^2p_3^2} } 
\nonumber\\
&&  +\dfrac{\sqrt{\pi}\;
\Gamma\left(\frac{d-2}{2} \right)}
{4\;\Gamma\left(\frac{d-1}{2}\right)}
\left( \dfrac{p_2^2 +p_3^2 -p_1^2}
{p_2^2p_3^2} \right) \; 
\left(-\dfrac{p_1^2}{4} \right)^{ \frac{d-4}{2}}\; 
\Fh21\Fz{1, \frac{d-2}{2}}{\frac{d-1}{2}} 
{- \dfrac{\lambda(p_1^2, p_2^2, p_3^2)}
{4p_2^2p_3^2} } 
\nonumber\\
&& \nonumber\\
&&  + 
\Big\{ (1,2,3) \rightarrow (2,3,1) \Big\} 
+\Big\{ (1,2,3) \rightarrow (3,1,2) \Big\}. 
\end{eqnarray}
Here $\lambda(x,y,z) =x^2+y^2+z^2-2xy-2xz-2yz$
is the K\"{a}ll\'{e}n function.
We remark that 
$p_i^2 \rightarrow p_i^2 + i\rho$
in this formula. With applying ($1.3.3.5$) in 
Ref.~\cite{Slater}, one can present $J_3$ as
\begin{eqnarray}
\label{j3massless} 
\dfrac{J_3}
{\Gamma\left( \frac{4-d}{2}\right) } &=&
\dfrac{2 }{(p_1^2-p_2^2 -p_3^2 )} 
 \left(\dfrac{p_1^2p_2^2p_3^2}
 {\lambda(p_1^2, p_2^2, p_3^2)} 
 \right)^{ \frac{d-4}{2}}
 \; \Fh21\Fz{1,\frac{1}{2}}{\frac{3}{2}} 
 { \dfrac{\lambda(p_1^2, p_2^2, p_3^2)}
 {(p_2^2+p_3^2-p_1^2)^2 } } \nonumber\\
&&  + \dfrac{\sqrt{\pi}\; 
 \Gamma\left(\frac{d-2}{2} \right)}
 {\Gamma\left(\frac{d-1}{2}\right)}
 \dfrac{
 \left(-p_1^2/4\right)^{ \frac{d-4}{2}}
}{p_2^2 +p_3^2 -p_1^2} 
 \; \Fh21\Fz{1, \frac{1}{2}}{\frac{d-1}{2}} 
 { \dfrac{\lambda(p_1^2, p_2^2, p_3^2)}
 {(p_2^2+p_3^2-p_1^2)^2} } \nonumber\\
&& \nonumber\\
&&  +
\Big\{ (1,2,3) \rightarrow (2,3,1) \Big\} 
+\Big\{ (1,2,3) \rightarrow (3,1,2) \Big\},
\end{eqnarray}
provided that 
$\left|\dfrac{\lambda(p_1^2, p_2^2, p_3^2)}
{(p_2^2+p_3^2-p_1^2)^2} \right|<1$ and 
$\mathcal{R}$e$(d-2)>0$. 
This equation is equivalent
to ($10$) in~\cite{Davydychev:1999mq}.
We note that we can
arrive to this result by 
inserting $J_2$ at $d+2s$ in 
(\ref{2point-fey-mij0}) into
(\ref{J3J31}) and taking the 
corresponding MB integrals.
\subsection{$R_{ij}=0$ }  
We consider the 
terms in $J_3$ with 
$R_{12} =0$ as an example. 
In this case, the terms 
$J_{231}^{(d)}$ and $J_{312}^{(d)}$ are 
given in the same form of (\ref{J123normal})
or (\ref{J123rotaion}). 
While the term $J_{123}^{(d)}$ 
is obtained
by performing analytic continuation the result in 
(\ref{J123rotaion}). In detail, one takes
the limit of $R_{12}\rightarrow 0$ in 
(\ref{J123rotaion}), we arrive at
\begin{eqnarray}
J_{123}^{(d)} 
&=&   
\frac{\left(\partial_3 R_3 \right)
\left(\partial_2 R_{12} \right) }{2(m_1^2 - R_3) } 
(m_1^2)^{\frac{d-4}{2} }
F_1 \left(1; \frac{4-d}{2}, 1; \frac{3}{2}; 
1, \dfrac{m_1^2}{m_1^2-R_3}  \right)   
+  (1\leftrightarrow 2). 
\end{eqnarray}
Using ($36$) in Ref.~\cite{schlosser}, 
the term $J_{123}^{(d)}$ simplifies to
\begin{eqnarray}
J_{123}^{(d)} 
&=&-\dfrac{\Gamma\left( \frac{d-3}{2}\right)}
{\Gamma\left( \frac{d-1}{2}\right)} \;
\left(\frac{\partial_3 R_3}{4R_3} \right) 
\left\{ 
\left(\partial_2 R_{12}\right)
(m_1^2)^{\frac{d-4}{2} } 
\Fh21\Fz{1,\frac{d-2}{2} }{\frac{d-1}{2} }
{ \dfrac{m_1^2}{R_3} } 
+  (1\leftrightarrow 2)      
\right\}, 
\end{eqnarray}
provided that 
$\mathcal{R}$e$\left(d-2\right)>0$ and 
$\left|m_i^2/R_3\right|<1$
for $i=1,2$. Taking $d\rightarrow4$, we have
\begin{eqnarray}
J_{123}^{(d=4)} 
&=&-\left(\frac{\partial_3 R_3}{2R_3} \right) 
\left\{ 
\left(\partial_2 R_{12}\right)
\Fh21\Fz{1,1 }{\frac{3}{2} }
{ \dfrac{m_1^2}{R_3} } 
+  (1\leftrightarrow 2)      
\right\}. 
\end{eqnarray}
\subsection{$R_{ij}=m_{i(j)}^2$ for $i,j=1,2,3$}
Next we consider
$R_{12} = m_1^2$ as an example. 
In this case, the 
terms $J_{231}^{(d)}, J_{312}^{d}$ are given 
by (\ref{J123normal}). Beside 
that, one verifies
\begin{eqnarray}
\partial_2 R_{12}
=0.  
\end{eqnarray}
As a result, we obtain
\begin{eqnarray}
J_{123}^{(d=4)}
&=&\dfrac{\left(\partial_3 R_3 \right)            
\left(\partial_1 R_{12}\right)}{2 (m_2^2 - R_3)} 
\Fh21\Fz{1,1 }{\frac{3}{2} }
{
\dfrac{m_1^2-m_2^2}{R_3 - m_2^2} },
\\
\label{rijmij}
J_{123}^{(d)}
&=&\dfrac{\left(\partial_3 R_3 \right)            
\left(\partial_1 R_{12}\right)}{2 (m_2^2 - R_3)} 
(m_2^2)^{\frac{d-4}{2} }
F_1 \left(1; \frac{4-d}{2}, 1; \frac{3}{2};
1-\frac{m_1^2}{m_2^2}, 
\dfrac{m_1^2-m_2^2}{R_3 - m_2^2}  \right),
\end{eqnarray}
provided that the amplitude of arguments of 
hypergeometric functions
appearing in this formula are less that $1$. 
For $R_{12} = m_1^2=m_2^2$, the function $F_1$ 
in (\ref{rijmij}) is equal $1$.  
The result reads
\begin{eqnarray}
\label{rijmii}
J_{123}^{(d)} 
&=&\dfrac{\left(\partial_3 R_3\right)}
{2(m^2- R_3) }
(m^2)^{\frac{d-4}{2} }.  
\end{eqnarray}
\subsection{$R_3=m_k^2$ for $k=1,2,3$} 
As an example, consider 
the terms of $J_3$ in (\ref{J3normal}) 
with $R_3 =m_1^2$. 
One verifies that
\begin{eqnarray}
\partial_1 R_3 = 1, \quad 
\partial_i R_3 = 0, 
\quad \text{for}\quad i=2, 3. 
\end{eqnarray}
As a result, $J_3$ 
is casted
into the form of
\begin{eqnarray}
\dfrac{J_3}
{\Gamma\left(\frac{4-d}{2}\right)}
= - J_{231}^{(d=4)}\; 
\left(R_3\right)^{\frac{d-4}{2}} 
+J_{231}^{(d)},
\end{eqnarray}
with $J_{231}^{(d)}$ taking 
the same form of 
(\ref{J123normal}) or (\ref{J123rotaion}). 
We take (\ref{J123rotaion}) as example for 
$J_{231}^{(d)}$. In detail, it takes
\begin{eqnarray}
\label{J231rotaion}
J_{231}^{(d)} &=&
\dfrac{\left(\partial_1 R_3\right) 
\left(\partial_3 R_{23} \right)}
{2(m_2^2-R_3)}  
(m_2^2)^{\frac{d-4}{2} }
F_1\left(1; \frac{4-d}{2}, 1; 
\frac{3}{2}; 1-\frac{R_{23} }{m_2^2}, 
\dfrac{R_{23}-m_2^2}
{R_3 - m_2^2}  \right)  
+  (2\leftrightarrow 3). 
\end{eqnarray}
Taking $d\rightarrow4$, the result reads
\begin{eqnarray}
\label{J231rotaion4}
J_{231}^{(d=4)} &=&
\dfrac{\left(\partial_1 R_3\right) 
\left(\partial_3 R_{23} \right)}
{2(m_2^2-R_3)}  
\Fh21\Fz{1,1 }{\frac{3}{2} }{
\dfrac{R_{23}-m_2^2}
{R_3 - m_2^2}  } 
+  (2\leftrightarrow 3). 
\end{eqnarray}
\subsection{$G_2\neq 0 $ and $R_3 = 0$} %
By setting $N=3$ in 
(\ref{mijkl0-sol}), one obtains
\begin{eqnarray}
\label{r1230}
J_3 
&=& \dfrac{1}{(d-4)} \sum\limits_{k=1}^3 
    \left(\frac{\partial_k Y_3}{G_2}\right) 
    {\bf k^{-} }  J_3(d-2; \{p_i^2\}, \{m_i^2\} ).
\end{eqnarray}
The resulting
of ${\bf k^{-} }  
J_3(d-2; \{p_i^2\}, \{m_i^2\})$ is 
$J_2$ in (\ref{J2F21-final}) or 
in (\ref{b7a}) with $d \rightarrow d-2$.
As an example, 
we take $J_2$ in 
(\ref{b7a}) at $d-2$. 
The result reads
\begin{eqnarray}
\label{R30}
\dfrac{J_3}{\Gamma(\frac{6-d}{2}) }
&=&\frac{1}{(4-d)}
\left(\frac{\partial_3 Y_3}{G_2}
\right)
\left\{ 
\left(\frac{\partial_2 R_{12} }
{R_{12} } \right)
(R_{12})^{\frac{d-4}{2}}
\Fh21\Fz{ \frac{6-d}{2}, \frac{1}{2} } 
{\frac{3}{2}}{1-\frac{m_1^2}{R_{12} } }
+
(1\leftrightarrow2)
\right\}
\nonumber\\
&& 
+ \Big\{(1,2,3) \leftrightarrow (2,3,1)\Big\} 
+  \Big\{(1,2,3) \leftrightarrow (3,1,2)\Big\}.
\end{eqnarray}
We can confirm (\ref{r1230}) again 
by using analytic continuation result of $J_3$
in (\ref{J123rotaion}). 
In fact, when $R_3 \rightarrow0$ 
the equation (\ref{J123rotaion}) becomes 
\begin{eqnarray}
 J^{(d)}_{123} &=&-\frac{1}{2}
\left(\frac{\partial_3 Y_3}{G_2}
\right)\left[
(\partial_2 R_{12}) (m_1^2)^{\frac{d-4}{2}}
F_1\left(1, \frac{4-d}{2}, 1,\frac{3}{2}, 
1-\frac{R_{12}}{m_1^2}, 
1-\frac{R_{12}}{m_1^2}  \right)
+
(1\leftrightarrow2)
\right] \\
 &=&-\frac{1}{2}
\left(\frac{\partial_3 Y_3}{G_2}
\right)\left\{
(\partial_2 R_{12}) (m_1^2)^{\frac{d-4}{2}}
\Fh21\Fz{ \frac{6-d}{2}, 1 } 
{\frac{3}{2}}{1-\frac{R_{12} }{m_1^2} }
+
(1\leftrightarrow2)
\right\} \\
&=&-\frac{1}{2}
\left(\frac{\partial_3 Y_3}{G_2}
\right)\left\{
\left(\frac{\partial_2 R_{12}}{R_{12}}
\right) (R_{12})^{\frac{d-6}{2}}
\Fh21\Fz{ \frac{6-d}{2}, 1 } 
{\frac{3}{2}}{1-\frac{m_1^2}{R_{12} } }
+
(1\leftrightarrow2)
\right\}. 
\label{R3rota0}
\end{eqnarray}
Plugging (\ref{R3rota0}) into 
(\ref{J3normal}), we arrive 
at (\ref{R30}).
\subsection{$R_3 = R_{ij}$ 
for $i,j =1,2,3$}          
We consider the case of $R_3 =R_{12}$
as an example. In this case, 
one verifies that  
\begin{eqnarray}
\partial_3 R_3 =0.
\end{eqnarray}
As a result, $J_3$ is
\begin{eqnarray}
\dfrac{J_3}
{\Gamma\left(\frac{4-d}{2}\right)}
&=& -J_{231}^{(d=4)}\;
\left(R_3\right)^{\frac{d-4}{2}} 
+J_{231}^{(d)} \nonumber\\
&& + \{(2,3,1) 
\leftrightarrow (3,1,2)\}. 
\end{eqnarray}
Here $J_{231}^{(d)}$ takes 
the same form as
(\ref{J123normal}) or (\ref{J123rotaion}).
We take (\ref{J123rotaion}) as example for 
$J_{231}^{(d)}$.
In detail, it takes
\begin{eqnarray}
J_{231}^{(d)} &=&
\dfrac{\left(\partial_1 R_3\right)       
\left(\partial_3  R_{23}\right) }
{ 2(m_2^2-R_3)}  
(m_2^2)^{\frac{d-4}{2} }
F_1\left(1; \frac{4-d}{2}, 1; 
\frac{3}{2}; 1-\frac{R_{23}}{m_2^2}, 
\dfrac{R_{23}-m_2^2}{R_3 - m_2^2}  \right)  
+  (2\leftrightarrow 3),\\
J_{231}^{(d=4)} &=&
\dfrac{\left(\partial_1 R_3\right)       
\left(\partial_3  R_{23}\right) }
{ 2(m_2^2-R_3)}  
\Fh21\Fz{1,1}{\frac{3}{2} }{
\dfrac{R_{23}-m_2^2}{R_3 - m_2^2}} 
+  (2\leftrightarrow 3). 
\end{eqnarray}
\subsection{$G_2 = 0$} 
Setting
$N=3$ in (\ref{GijkN0}), 
the result reads
\begin{eqnarray}
\label{g123J30}
J_3 
= -\frac{1}{2}\sum\limits_{k=1}^3 
\left(\frac{\partial_k Y_3 } {Y_3 } \right) 
{\bf k^{-} }  J_3(d; \{p_i^2\}, \{m_i^2\} ).
\end{eqnarray}
This equation is equivalent to
($46$) in Ref.~\cite{Devaraj:1997es}. The term 
${\bf k^{-} }  J_3(d; \{p_i^2\}, \{m_i^2\})$
corresponds to
$J_2$ in (\ref{b7a}) 
as an example. 
We obtain
\begin{eqnarray}
\dfrac{J_3}{\Gamma(\frac{4-d}{2}) }
&=&-\left(\frac{\partial_3 Y_3}{2Y_3}
\right)
\left\{ 
\left(\frac{\partial_2 Y_{12} }
{Y_{12} } \right)
(m_1^2)^{\frac{d-2}{2}}
\Fh21\Fz{ \frac{d-1}{2}, 1 } 
{\frac{3}{2}}{1-\frac{m_1^2}{R_{12} } }
+
(1\leftrightarrow2)
\right\}
\nonumber\\
&& 
+ \Big\{(1,2,3) \leftrightarrow (2,3,1)\Big\} 
+  \Big\{(1,2,3) \leftrightarrow (3,1,2)\Big\}.
\end{eqnarray}
\subsection{$G_{1(ij)} = 0$ for $i,j=1,2,3$} 
$G_{1(ij)}$ are the Gram 
determinants of two-point functions 
which are obtained by shrinking a
propagator $k\neq i,j$
in the three-point ones. Taking $G_{1(12)}=0$
as an example, the term 
$J_{123}^{(d)}$ is evaluated
as follows. We put $J_2$ in
(\ref{J2G20}) into (\ref{J3J31}). Taking 
the corresponding MB integrations, the results 
read as form of
(\ref{J3normal}) with
\begin{eqnarray}
J_{123}^{(d=4)} &=&  - 
\left(\dfrac{\partial_3 R_3}{R_3} \right) 
\dfrac{m_1^2}{ m_1^2-m_2^2} 
\Fh21\Fz{ 1,1}{2}
{\frac{m_1^2}{R_3}}      
+(1 \leftrightarrow 2), \\
J_{123}^{(d)} &=&  
- \dfrac{\Gamma\left(\frac{d-2}{2}\right)}
{2\Gamma\left(\frac{d}{2}\right)} 
\left(\dfrac{\partial_3 R_3}{R_3} \right)   
\dfrac{(m_1^2)^{\frac{d-2}{2}}}{ m_1^2-m_2^2} 
\Fh21\Fz{ \frac{d-2}{2} ,1}{\frac{d}{2}}
{\frac{m_1^2}{R_3}}      
+(1 \leftrightarrow 2).
\end{eqnarray}
It is valid under the conditions 
that the arguments of 
the hypergeometric fuctions appearing in 
this formula are less that $1$ and 
$\mathcal{R}$e$(d-2)>0$.
\subsection{Cross check with other papers}
We consider $p_2^2=p_3^2=0;p_1^2\neq 0,
m_1^2=m_3^2=0$ and $m_2^2\neq 0$ as an 
example~\cite{Abreu:2015zaa,Phan:2019qee}. 
We confirm that 
\begin{eqnarray}
R_3 = \dfrac{m_2^2 (p_1^2 + m_2^2)}{p_1^2}, 
\dfrac{\partial_1 R_3}{R_3}
=\dfrac{\partial_3 R_3}{R_3}
= -\dfrac{1}{p_1^2 +m_2^2},         
\dfrac{\partial_2 R_3}{R_3}
= \dfrac{2m_2^2 +p_1^2}{m_2^2(m_2^2 +p_1^2)}.             
\end{eqnarray}
The $J_3$ in (\ref{J3J31}) becomes
\begin{eqnarray}
\label{j3crosscheck}
J_3 &=&
-\dfrac{1}{2\pi i} 
\int\limits_{-i\infty}^{+i\infty}\;ds \; 
\dfrac{\Gamma(-s)\;\Gamma(s+1)
\Gamma\Big(\frac{d-2}{2}+s\Big)
\Gamma\Big(\frac{4-d}{2}-s\Big)}
{2\;\Gamma(\frac{d-2}{2})} 
\left(\dfrac{1}{R_3 } \right)^s  
\\
&&
\times \Bigg\{
\dfrac{\sqrt{\pi}}{2}
\dfrac{2m_2^2 +p_1^2}{m_2^2(m_2^2 +p_1^2)}
\dfrac{\Gamma\left(\frac{d-2}{2}+s\right) }
{ \Gamma\left(\frac{d-1}{2}+s\right)}\;
\left(\dfrac{-p_1^2 }{4}\right)^{ \frac{d-4}{2}+s}
-\frac{2}{p_1^2 +m_2^2} 
\dfrac{\Gamma\left(\frac{d-2}{2}+s\right) }
{ \Gamma\left(\frac{d}{2}+s\right)}\; 
(m_2^2)^{\frac{d-4}{2}+s}
\Bigg\}.
\nonumber
\end{eqnarray}
We note that the first term
in curly bracket of (\ref{j3crosscheck}) is $J_2$ 
in the case of (\ref{2point-fey-mij0}) with 
$d$ shifted to $d+2s$. While the second
term in curly bracket
of (\ref{j3crosscheck}) is corresponding to 
$J_2$ in (\ref{J2G20}) at 
$d+2s$ (and with a massless internal line). 
In the following 
we perform the contour integration 
of (\ref{j3crosscheck}) starting with the 
second contour integral:
\begin{eqnarray}
J_3^{(1)} 
&=& 
-\dfrac{ (m_2^2)^{\frac{d}{2}-2}\; }{2\pi i} 
\int\limits_{-i\infty}^{+i\infty}\;ds \; 
\dfrac{\Gamma(-s)\;\Gamma(s+1)
\Gamma\Big(\frac{d-2}{2}+s\Big)^2 
\Gamma\Big(\frac{4-d}{2}-s\Big)}
{2\;\Gamma(\frac{d-2}{2} )\;  
\Gamma\left(\frac{d}{2}+s\right)} 
\left(\dfrac{m_2^2}{R_3 } \right)^s.
\end{eqnarray}
By closing the integration  contour 
to the right, the residue contributions 
at the poles of $\Gamma(-s)$ and 
$\Gamma\left(\frac{4-d}{2} - s\right)$ are 
calculated. For the first sequence poles, 
the result reads
\begin{eqnarray}
J_3^{(1,a)}
&=& 
-\dfrac{\Gamma\left( 2-\frac{d}{2}\right) 
\Gamma\left( \frac{d}{2}-1\right)}
{ 2 \Gamma\left( \frac{d}{2}\right) } 
(m_2^2)^{\frac{d}{2}-2}\;
\Fh21\Fz{1, \frac{d-2}{2}}{\frac{d}{2}} 
{ \dfrac{p_1^2}{p_1^2 +m_2^2}}                                     \\
&=& 
-\dfrac{\Gamma\left( 2-\frac{d}{2}\right)
\Gamma\left( \frac{d}{2}-1\right)}
{2 \Gamma\left( \frac{d}{2}\right) }
\dfrac{p_1^2+m_2^2}{m_2^2}\;
(m_2^2)^{\frac{d}{2}-2}\;
\Fh21\Fz{1, 1}{\frac{d}{2}} 
{ -\dfrac{p_1^2}{m_2^2}}.             
\end{eqnarray}
For the second sequence poles, 
we arrive at
\begin{eqnarray}
J_3^{(1,b)}  
&=&
- \dfrac{\Gamma\left( \frac{d}{2}-2\right)
\Gamma\left( 3-\frac{d}{2}\right)}
{2\Gamma\left(\frac{d}{2}-1 \right)} 
\; (R_3)^{\frac{d}{2} - 2} \;
\Fh21\Fz{1,1}{2} 
{\dfrac{p_1^2}{p_1^2 +m_2^2}}  \\
&=&-\dfrac{\Gamma\left( \frac{d}{2}-2\right)
\Gamma\left( 3-\frac{d}{2}\right)}
{2\Gamma\left(\frac{d}{2}-1 \right)} 
\; (R_3)^{\frac{d}{2} - 2} \;
\dfrac{p_1^2+m_2^2}{m_2^2} 
\Fh21\Fz{1,1}{2} {- \dfrac{p_1^2}{m_2^2}}.                              \nonumber\\
\end{eqnarray}
Second type of MB integral 
is considered 
\begin{eqnarray}
J_3^{(2)}&=&
- \dfrac{ 1 }{2\pi i} 
\int\limits_{-i\infty}^{+i\infty}\;ds \; 
\dfrac{\sqrt{\pi}}{2}\;
\dfrac{\Gamma(-s)\;\Gamma(s+1) 
\Gamma\Big(\frac{d-2}{2}+s\Big)^2 
\Gamma\Big(\frac{4-d}{2}-s\Big)}
{2\;\Gamma(\frac{d-2}{2})\;  
\Gamma\left(\frac{d-1}{2}+s\right)}  
\left(\dfrac{1}{R_3 } \right)^s \;  
\left( \dfrac{-p_1^2}{4}
\right)^{\frac{d}{2}-2+s}.
\nonumber\\
\end{eqnarray}
For the first sequence poles
of $\Gamma(-s)$, the result is
\begin{eqnarray}
J_3^{(2,a)} 
&=&
-\dfrac{\sqrt{\pi}\; 
\Gamma\left( \frac{d}{2}-1\right) 
\Gamma\left( 2-\frac{d}{2}\right)}
{4\;\Gamma\left(\frac{d-1}{2} \right)} 
\left( \dfrac{-p_1^2 }{4}
\right)^{\frac{d}{2} - 2}
\Fh21\Fz{1,\frac{d-2}{2}}{\frac{d-1}{2}} 
{\dfrac{-(p_1^2)^2}{4m_2^2(p_1^2 +m_2^2)}}.  
\nonumber
\end{eqnarray}
Applying transforms for Gauss 
hypergeometric function which are
(see $(1.8.10)$ in \cite{Slater} 
for the first relations 
and  page $49$, 
\cite{berndt} for the later case)
\begin{eqnarray}
\Fh21\Fz{a, b}{c}{z} 
  &=& (1-z)^{c-b-a} \Fh21\Fz{c-a, c-b}{c}{z},                        \\ 
\Fh21\Fz{a, b}{2b}{z} 
&=& (1-z)^{-a/2} \Fh21\Fz{\frac{a}{2}, 
b-\frac{a}{2}}{b +\frac{1}{2}}
{\dfrac{z^2}{4(z-1)}},
\end{eqnarray}
one obtains
\begin{eqnarray}
J_3^{(2,a)}
&=& -\dfrac{\sqrt{\pi}}{2}\;   
\dfrac{\Gamma\left( \frac{d}{2}-1\right) 
\Gamma\left( 2-\frac{d}{2}\right)}
{\Gamma\left(\frac{d-1}{2} \right)}\;  
\left( \dfrac{-p_1^2}{4}
\right)^{\frac{d}{2} - 2}
\dfrac{(p_1^2+m_2^2)}{(p_1^2 + 2m_2^2)} 
 \Fh21\Fz{1,\frac{d-2}{2}}{d-2} 
 {- \dfrac{p_1^2 }{m_2^2}}. 
\end{eqnarray}
Taking into account the residue at the 
poles $\Gamma\left(\frac{4-d}{2}-s \right)$, 
we get
\begin{eqnarray}
J_3^{(2,b)}  
=-\dfrac{\Gamma\left( \frac{d}{2}-2\right) 
\Gamma\left( 3-\frac{d}{2}\right)}
{2\Gamma\left(\frac{d}{2} -1\right)}\;  
\left[ \dfrac{4 m_2^2 (p_1^2 +m_2^2)}
{ (2m_2^2 +p_1^2)^2} \right]
\; R_3^{\frac{d}{2} -2} \; 
\Fh21\Fz{1,\frac{1}{2}}{\frac{3}{2}}
{\dfrac{(p_1^2)^2}{(2m_2^2 +p_1^2)^2} }.                                                       
\end{eqnarray}
Using the relation 
(see Eq.~$(3.1.7)$ in \cite{roy}) 
\begin{eqnarray}
\Fh21\Fz{a, b}{2b}{z} 
&=& \left( 1-\frac{z}{2} \right)^{-a} 
\Fh21\Fz{\frac{a}{2}, \frac{a}{2} 
+ \frac{1}{2} }{b +\frac{1}{2}} 
{\left(\frac{z}{2-z} \right)^2},
\end{eqnarray}
one gets
\begin{eqnarray}
J_3^{(2,b)} 
&=&  
-\dfrac{\Gamma\left( \frac{d}{2}-2\right) 
\Gamma\left( 3-\frac{d}{2}\right)}
{\Gamma\left(\frac{d}{2} -1\right)}
\;\dfrac{ (p_1^2 +m_2^2)}{ 2m_2^2 +p_1^2}
\;  R_3^{\frac{d}{2} -2}\; 
\Fh21\Fz{1,1}{2} {-\dfrac{p_1^2}{m_2^2} }.    
\end{eqnarray}
Combining all the terms, $J_3$ reads
\begin{eqnarray}
J_3 
&=&
\dfrac{\Gamma\left( 2-\frac{d}{2}\right) 
\Gamma\left( \frac{d}{2}-1\right)}
{ \Gamma\left( \frac{d}{2}\right) }
(m_2^2)^{\frac{d}{2}-3} 
\Fh21\Fz{1, 1}{\frac{d}{2}} 
{-\dfrac{p_1^2}{m_2^2}}  \\
& & 
-\dfrac{\sqrt{\pi}}{2}\;   
\dfrac{\Gamma\left(\frac{d}{2}-1\right) 
\Gamma\left( 2-\frac{d}{2}\right)}
{\Gamma\left(\frac{d-1}{2} \right)\; m_2^2}\;  
\left( \dfrac{-p_1^2}{4}
\right)^{\frac{d}{2}- 2}
\Fh21\Fz{1,\frac{d-2}{2}}{d-2} 
{-\dfrac{p_1^2 }{m_2^2}}                                                   \nonumber
\end{eqnarray}
provided that 
$\left|p_1^2/m_2^2\right|<1$ and 
$\mathcal{R}$e$(d-2)>0$. It agrees 
with Eq.~($B2$) in~\cite{Abreu:2015zaa}
and ($22$) in \cite{Phan:2019qee}.
\section{One-loop four-point functions}
The master equation for $J_4$ is obtained 
from (\ref{JNJN1}) with $N=4$, 
\begin{eqnarray}
\label{J4J3sect4}
J_4 \equiv
J_4 (d;\{p_i^2, s, t\},\{m_i^2\}) 
&=& -\dfrac{1}{2\pi i} 
\int\limits_{-i\infty}^{+i\infty}ds \; 
\dfrac{\Gamma(-s)\; 
\Gamma(\frac{d-3}{2}+s) \Gamma(s+1) }
{ 2\Gamma(\frac{d-3}{2}) } 
\left(\frac{1}{R_4}\right)^s                
\times\\
&&\hspace{1.0cm}\times    
\sum\limits_{k=1}^4 
\left(\frac{\partial_k R_4 }{R_4}\right) 
\; {\bf k}^-J_4 (d+2s;\{p_i^2, s, t\},\{m_i^2\}).  
\nonumber
\end{eqnarray}
We substitute the
analytic solution for 
$J_3(d+2s; \{p_i^2\},\{m_i^2\}) $ 
in~(\ref{J3normal}) 
into~(\ref{J4J3sect4}) and
take the contour integrals 
in (\ref{J4J3sect4}). With the help 
of MB integrations in
(\ref{MBFS}, \ref{MBFS-b}) in 
appendix $C$, a compact expression
for $J_4$ can be derived and expressed 
as follows:
\begin{eqnarray}
\label{J4normal}
\dfrac{J_4}
{ \Gamma\left( \frac{4-d}{2} \right) } 
&=& - J_{1234}^{(d=4)}  
\left(R_4\right)^{\frac{d-4}{2}}
+ J_{1234}^{(d)}\nonumber\\
&& + \Big\{(1,2,3,4) 
\leftrightarrow (2,3,4,1)\Big \} \\
&&\hspace{0cm}+ 
\Big\{(1,2,3,4) \leftrightarrow (3,4,1,2) \Big\}
\nonumber\\
&&+ \Big\{(1,2,3,4) 
\leftrightarrow (4,1,2,3) \Big\}
\nonumber
\end{eqnarray}
with
\begin{eqnarray}
\label{j1234normal}
&& \hspace{-1.5cm}
J_{1234}^{(d)}
= -\left(\frac{\partial_4 R_4 }
{2 R_4}\right) \; J_{123}^{(d=4)}
\;  \left( R_{123}\right)^{\frac{d-4}{2}} 
\Fh21\Fz{ \frac{d-3}{2}, 1}{\frac{d-2}{2} }
{\frac{R_{123}}{R_4} }
\\
&& + \dfrac{\sqrt{\pi} 
\Gamma\left(\frac{d-2}{2}\right) }
{\Gamma(\frac{d-1}{2}) }
\left(
\frac{\partial_4 R_4 }{R_4} 
\right)
\left(
\frac{\partial_3 R_{123} }{R_{123}}
\right)
\left[\dfrac{\partial_2 R_{12} }
{\sqrt{1-m_1^2/R_{12} }}
+ \dfrac{\partial_1 R_{12} }
{\sqrt{1-m_2^2/R_{12} }} \right] 
\times \nonumber\\
&&   \hspace{4cm}          
 \times
 \frac{~~~\left(R_{12}\right)^{\frac{d-4}{2}}}
 {\sqrt{1-R_{12}/R_{123}}}
 F_1\left(\frac{d-3}{2}; 1, \frac{1}{2}; \frac{d-1}{2}; 
  \frac{R_{12} }{R_4}, 
  \frac{R_{12} }{R_{123}}  \right) 
\nonumber\\
&&\hspace{0cm}
- \dfrac{\Gamma\left(\frac{d-2}{2}\right) }
{8\;\Gamma(\frac{d}{2}) }
\left(
\frac{\partial_4 R_4 }{R_4} 
\right)\Bigg[
\frac{\partial_3 R_{123} }{(R_{123} -m_1^2)} 
\frac{\partial_2 R_{12} }
{(R_{12} -m_1^2)} (m_1^2)^{\frac{d-2}{2}}
\times \nonumber\\
&&   \hspace{0.2cm} \times 
F_S\left( \frac{d-3}{2},1,1; 1,1,\frac{1}{2}; 
               \frac{d}{2},\frac{d}{2},\frac{d}{2};
               \frac{m_1^2}{R_4},
               \frac{m_1^2}{m_1^2-R_{123}}, 
               \frac{m_1^2}{m_1^2-R_{12} }                        
               \right)  
     +  (1\leftrightarrow 2)     
\Bigg]  
\nonumber\\
&& \nonumber\\
&&\hspace{-0cm} + 
\Big\{(1,2,3)\leftrightarrow (2,3,1)\Big\} 
\quad 
+
\quad
\Big\{(1,2,3)\leftrightarrow (3,1,2)
\Big\}
. \nonumber
\end{eqnarray}
Where 
$J_{123}^{(d=4)}, \cdots$ are given by 
(\ref{J123normal4}). It is important 
that this representation is 
valid under the conditions 
that $\mathcal{R}$e$\left(d-3\right)>0$ 
and the absolute values
of arguments of 
hypergeometric functions
{are smaller than one}. 
If the absolute value of these arguments
{are larger than one,} 
we have to perform  analytic continuations 
for the Gauss hypergeometric and 
Appell $F_1$ functions, cf.~\cite{Slater,olsson}. 
{Further, the Saran function 
$F_S$ may be expressed} 
by a Mellin-Barnes 
representation, 
or Euler integrals in this case.
The result for $J_4$ has been shown 
in~\cite{Phan:2018cnz}.
There are two important points 
we would like to emphasize in this paper as follows. 
(i) Ref.~\cite{Fleischer:2003rm} have not
shown conditions for the boundary term in $(100)$. 
(ii) $J_4$ is constructed
from $J_3$ for arbitrary 
kinematics. However, the boundary
term for $J_3$ for 
general kinematics have not been
provided in \cite{Fleischer:2003rm}, as mentioned
in the 
previous section and in \cite{Bluemlein:2017rbi}. 
Subsequently, the first term in $(99)$ 
of \cite{Fleischer:2003rm} 
is only valid in special kinematic regions. Therefore, 
the solution in ($98$) of Ref.~\cite{Fleischer:2003rm} 
may not be considered as a 
complete solution for $J_4$. 
In contrast to \cite{Fleischer:2003rm}, 
we provide a complete solution for $J_4$ in this article.
\subsection{Massless internal lines}%
We are going to take $m_i^2 \rightarrow 0$ 
for $i=1,2,3,4$. The terms 
related to $F_S$ vanish. 
Therefore, in the massless case
the result reads 
\begin{eqnarray}
\label{J1234massless}
J_{1234}^{(d)}
&=&  \left(\frac{\partial_4 R_4 }{2 R_4 }\right) 
     \Big|_{m_i^2\rightarrow0}\; 
     \left( 
\frac{\partial_3 R_{123} }{2R_{123} } 
\right)
\Big|_{m_i^2\rightarrow0}
\Fh21\Fz{1, 1}{\frac{3}{2}}
{ \dfrac{R_{12} }{R_{123} } }     
     \; \left(R_{123}\right)^{\frac{d-4}{2}} 
     \Fh21\Fz{ \frac{d-3}{2}, 1}{\frac{d-2}{2}}
     {\frac{R_{123}}{R_4} }
\nonumber\\
& &\hspace{0cm} 
+\dfrac{\sqrt{\pi}
\Gamma\left(\frac{d-2}{2}\right) }
{2\Gamma(\frac{d-1}{2}) }
\left( \frac{\partial_4 R_4 }{2 R_4} \right)
\Big|_{m_i^2\rightarrow0}
\left(\frac{\partial_3 R_{123} }
{2 R_{123}} \right)
\Big|_{m_i^2\rightarrow0}
\nonumber\\
&&   \hspace{0.2cm}\times  
\frac{\left(R_{12}\right)^{\frac{d-4}{2} }}
{\sqrt{1-R_{12}/R_{123}}}
F_1\left( \frac{d-3}{2}; 1, \frac{1}{2}; 
\frac{d-1}{2}; \frac{R_{12} }{R_4},
\frac{R_{12} }{R_{123}} 
\right) \nonumber\\
&& \nonumber\\
&&\hspace{0cm}  +
\{(1,2,3)\leftrightarrow (2,3,1)\} 
\quad 
+
\quad
\{(1,2,3)\leftrightarrow (3,1,2)
\}, 
\end{eqnarray}
provided that $\mathcal{R}$e$\Big(d-3\Big)>0$ 
and {that} the absolute values
of arguments of the hypergeometric 
functions {are smaller than one}.
Taking $d\rightarrow4$, we have
\begin{eqnarray}
\label{J1234massless4}
J_{1234}^{(d=4)}
&=&  
\left(\frac{\partial_4 R_4 }{2\;R_4 }\right) 
\Big|_{m_i^2\rightarrow0}\; 
\left( 
\frac{\partial_3 R_{123} }{2R_{123} } 
\right)
\Big|_{m_i^2\rightarrow0}
\Fh21\Fz{1, 1}{\frac{3}{2}}
{ \dfrac{R_{12} }{R_{123} } }     
\Fh21\Fz{ \frac{1}{2}, 1}{1}
{\frac{R_{123}}{R_4} }
\nonumber\\
& &\hspace{0cm} 
+ \left( \frac{\partial_4 R_4 }{2 R_4} \right)
\Big|_{m_i^2\rightarrow0}\;
\left(\frac{\partial_3 R_{123} }
{2 R_{123}} \right)
\Big|_{m_i^2\rightarrow0}
\left(\frac{R_{123}}{R_{123}-R_{12}}\right)
\Fh21\Fz{\frac{1}{2}, 1}{\frac{3}{2}}
{ \dfrac{R_{12}(R_{123} -R_4) }
{R_4(R_{123}-R_{12}) } }     
\nonumber\\
&& \nonumber\\
&&\hspace{0cm}  +
\{(1,2,3)\leftrightarrow (2,3,1)\} 
\quad 
+
\quad
\{(1,2,3)\leftrightarrow (3,1,2)
\}, 
\end{eqnarray}
This is a new result for $J_4$ 
in the massless case at general $d$.
We are going to consider the special cases for 
$J_4$ in the following subsections.
\subsection{$R_4 = 0$} 
From (\ref{mijkl0-sol}), we set $N=4$ and get
\begin{eqnarray}
\label{J4m12340}
J_4 = \frac{1}{d-5} \sum\limits_{k=1}^4
\left(\frac{\partial_k Y_4}
{G_3}\right) 
{\bf k^{-} } J_4(d-2;\{p_i^2, s, t\}, \{m_i^2\}).
\end{eqnarray}
The term ${\bf k^{-} } 
J_4(d-2;\{p_i^2, s, t\}, \{m_i^2\})$
is given by $J_3$ in (\ref{J3normal}) 
with $d\rightarrow d-2$.
\subsection{$R_4 =       
R_{ijk}$ for $i,j,k =1,2,3,4$} 
As an example, we consider 
the case $R_4 = R_{123}$.
In this case, we verify that 
\begin{eqnarray}
\left(\partial_4 R_4\right) = 0.  
\end{eqnarray}
As a result, the terms $J_{1234}^{(d)}$
vanish, other terms 
in (\ref{J4normal}) are of
the same form in (\ref{j1234normal}). 
\subsection{$R_4 =             
R_{ij}$ for  $i,j = 1,2,3,4$ } 
For example, the terms of $J_4$ in 
(\ref{J4normal})
meet the condition $R_4 = R_{12}$.
Because that $R_2$ depends only the 
internal masses $m_1^2, m_2^2$, one has
\begin{eqnarray}
\partial_i R_4 = \partial_i R_2 = 0, 
\quad \text{for} \quad i=3,4. 
\end{eqnarray}
As a matter of this fact, only two terms 
$(2,3,4,1)$ and $(3,4,1,2)$
in (\ref{J4normal}) contribute to $J_4$.
\subsection{$R_4 = m_k^2$
\; for $k=1,2,3,4$}         
For example, one considers 
the terms of $J_4$ in (\ref{J4normal}) 
having $R_4 =m_1^2$. One verifies that
\begin{eqnarray}
\partial_i R_4 = 0 \quad \text{for} 
\quad i=2, 3, 4. 
\end{eqnarray}
As a result, only the term $(2,3,4,1)$ in 
(\ref{J4normal}) contributes
to $J_4$. 
\subsection{$R_{ijk} = 0$ 
 for $i,j,k=1,2,3,4$}
We assume that $J_4$ in (\ref{J4normal}) 
contains $R_{123} =0$ as an example.
The term $(1,2,3,4)$ in (\ref{j1234normal}) 
with $R_{123}=0$ is evaluated
by applying the same previous 
procedure. The result reads
\begin{eqnarray}
J_{1234}^{(d)} 
&=& -\dfrac{\sqrt{\pi} }
{\Gamma(\frac{d-3}{2}) \Gamma(\frac{d-2}{2}) }
\left( \frac{\partial_4 R_4 }{R_4} \right)
\left( \frac{\partial_3 Y_{123} }
{G_{12}} \right)\;  
\times \\
&&\hspace{0cm}\times \left[
\dfrac{\partial_2 R_{12} }
{\sqrt{1-m_1^2/R_{12} }} 
+ \dfrac{\partial_1 R_{12} }
{\sqrt{1-m_2^2/R_{12} }}  
\right]\; \left(R_{12} \right)^{\frac{d-6}{2}} 
\Fh21\Fz{1, \frac{d-4}{2}} {\frac{d-2}{2}}
{ \dfrac{R_{12} }{R_4} }  
\nonumber\\
&&  + \dfrac{1}
{2\;\Gamma^2(\frac{d-2}{2}) }
\left(\frac{\partial_4 R_4 }{R_4} \right)
\left( \frac{\partial_3 Y_{123} }
{G_{12}} \right)
\times \nonumber\\
&&\hspace{0cm}
\times\left[
\left(\frac{\partial_2 R_{12} }{ 
R_{12} } \right)
\dfrac{~~(m_1^2)^{\frac{d-4}{2}} }
{ \sqrt{1- m_1^2/R_{12} }}  F^{1; 2;1}_{1;1;0}
\left(\begin{matrix}
\frac{d-4}{2};~~~
\frac{d-3}{2}, ~~~1; ~~~\frac{1}{2};\\
\frac{d-2}{2};~~~\frac{d-2}{2}; -;
\end{matrix}~~~
\frac{m_1^2}{R_4}, 
\frac{m_1^2}{R_{12} }  \right)
  +  (1\leftrightarrow 2)      
\right]\nonumber\\
&&                                  
\nonumber\\
&&  \hspace{0cm} +
\{(1,2,3)\leftrightarrow (2,3,1)\} 
\quad
+
\quad
\{(1,2,3)\leftrightarrow (3,1,2)
\}.   
\nonumber
\end{eqnarray}

Where $F^{1; 2;1}_{1;1;0}$ is 
Kamp$\acute{\text{e}}$ de 
F$\acute{\text{e}}$riet
\cite{KampedeFeriet:37}
(see appendix
$B$ for more detail).
We also refer to 
\cite{Ananthanarayan:2020acj}
which analytic continuations for 
a class of the Kamp$\acute{\text{e}}$ de 
F$\acute{\text{e}}$riet functions have been 
studied. This representation is valid
if the amplitude of arguments of 
these hypergeometric functions 
are less than $1$ and 
$\mathcal{R}$e$(d-4)>0$. 
In the massless
case, one has 
\begin{eqnarray}
 J_{1234}^{(d)} 
&=&\dfrac{\sqrt{\pi} }
{\Gamma(\frac{d-3}{2}) 
\Gamma(\frac{d-2}{2}) }
\left(
\frac{\partial_4 R_4 }{R_4}\right)
\left( \frac{\partial_3 Y_{123} }
{G_{12}} \right)
\left(R_{12} \right)^{\frac{d-6}{2}}
\Fh21\Fz{1, \frac{d-4}{2}} {\frac{d-2}{2}}
{ \dfrac{R_{12} }{R_4} } 
\nonumber \\
&&                                  
\nonumber\\
&&  \hspace{0cm} 
+
\{(1,2,3)\leftrightarrow (2,3,1)\} 
\quad
+
\quad
\{(1,2,3)\leftrightarrow (3,1,2)
\}.
\end{eqnarray} 
\subsection{$R_{ijk} = R_{ij}$ 
for $i,j, k=1,2,3,4$} 
We examine the terms of $J_4$ in 
(\ref{J4normal}) having $R_{123} = R_2$.
We check that  
\begin{eqnarray}
\partial_3 R_{123} = 0.
\end{eqnarray}
As a result, the terms $(2,3,4,1), (3,4,1,2)$ 
and $(4,1,2,3)$ of $J_4$ get
the same formula for $J_{1234}^{(d)}$ 
provided in (\ref{j1234normal}). 
The terms $(1,2,3)$ of $J_{1234}^{(d)}$ in 
(\ref{j1234normal}) are  
vanished. 
\subsection{$R_{ijk} = m_{i,(j,k)}^2$ 
for $i,j,k = 1,2,3,4$}
Assuming the terms of $J_4$ in 
(\ref{J4normal}) with
$R_{123} = m_1^2$, we check that  
\begin{eqnarray}
\partial_1 R_{123} = 1,
 \quad \text{and} \quad 
 \partial_i R_{123} = 0, 
 \quad \text{for} \quad i=2, 3. 
\end{eqnarray}
The terms $J_{2341}^{(d)}, J_{3412}^{(d)}$ 
and $J_{4123}^{(d)}$ get the same 
formula as (\ref{j1234normal}). 
The term $(1,2,3)$ and 
$(3,1,2)$ of  $J_{1234}^{(d)}$ in 
(\ref{j1234normal}) vanish,
only the $(2,3,1)$-term contribute to 
$J_{1234}^{(d)}$. 
\subsection{$R_{ij} = 0$ for $i,j=1,2,3,4$}
We consider that $J_4$ in (\ref{J4normal}) 
with
$R_{12} =0$. In this case, the terms 
$J_{2341}^{(d)}$ and $J_{3412}^{(d)}$
are unchanged. The terms $J_{1234}^{(d)}$ 
and $J_{4123}^{(d)}$ with
$R_{12} = 0$ are evaluated
again by applying same previous procedure. Taking 
$J_{1234}^{(d)}$ as an example.
The result reads
\begin{eqnarray}
J_{1234}^{(d)} 
&=&- \left( 
\frac{\partial_4 R_4 }{R_4} \right)
\; J_{123}^{(d=4)}
\;(R_{123})^{\frac{d-4}{2}}
\Fh21\Fz{1,\frac{d-3}{2} }{\frac{d-2}{2} }
{ \dfrac{R_{123}}{R_4} }     
\nonumber\\
&&+ \dfrac{ \Gamma\left(\frac{d-3}{2}\right) }
{4\Gamma(\frac{d-1}{2}) }
\left( \frac{\partial_4 R_4 }{R_4}
\right)
\left( \frac{\partial_3 R_{123} }{R_{123} 
}\right) 
\times     \\
&&\hspace{0cm} \times 
\Bigg[ 
\left(
\dfrac{\partial_2 Y_{12}}{G_{12}}
\right)
\frac{(m_1^2)^{\frac{d-4}{2}}}
{\sqrt{ 1-m_1^2/R_{123} }}    
F^{1;\;2;\;1}_{1;\;1;\;0}
\left(
\begin{matrix}
 \frac{d-3}{2};~~~\frac{d-3}{2},~~~1;
 ~~~\frac{1}{2};\\
 \frac{d-1}{2},~~~\frac{d-2}{2};-;
\end{matrix}~~~
\dfrac{m_1^2}{R_4}, ~~~
\dfrac{m_1^2}{R_{123}} 
\right)
+  (1\leftrightarrow 2)  
\Bigg]  \nonumber\\
&& \nonumber\\
&&\hspace{-0cm}
+\Bigg\{ 
-\left(\frac{\partial_4 R_4 }
{2 R_4}\right) \; J_{231}^{(d=4)}
\;  \left( R_{231}\right)^{\frac{d-4}{2}} 
\Fh21\Fz{ \frac{d-3}{2}, 1}{\frac{d-2}{2} }
{\frac{R_{231}}{R_4} }
\nonumber\\
&& \hspace{0.7cm}+ \dfrac{\sqrt{\pi} 
\Gamma\left(\frac{d-2}{2}\right) }
{\Gamma(\frac{d-1}{2}) }
\left(
\frac{\partial_4 R_4 }{R_4} 
\right)
\left(
\frac{\partial_1 R_{231} }{R_{231}}
\right)
\left[\dfrac{\partial_3 R_{23} }
{\sqrt{1-m_2^2/R_{23} }}
+ \dfrac{\partial_1 R_{23} }
{\sqrt{1-m_2^2/R_{23} }} \right] 
\times \nonumber\\
&&   \hspace{4cm}          
 \times
 \frac{~~~\left(R_{23}\right)^{\frac{d-4}{2}}}
 {\sqrt{1-R_{23}/R_{231}}}
 F_1\left(\frac{d-3}{2}; 1, \frac{1}{2}; \frac{d-1}{2}; 
  \frac{R_{23} }{R_4}, 
  \frac{R_{23} }{R_{231}}  \right) 
\nonumber\\
&&
\hspace{0.7cm}
-\dfrac{\Gamma\left(\frac{d-2}{2}\right) }
{8\;\Gamma(\frac{d}{2}) }
\left(
\frac{\partial_4 R_4 }{R_4} 
\right)\Bigg[
\frac{\partial_1 R_{231} }{(R_{231} -m_2^2)} 
\frac{\partial_3 R_{23} }
{(R_{23} -m_2^2)} (m_2^2)^{\frac{d-2}{2}}
\times \nonumber\\
&&   \hspace{0.8cm} \times 
F_S\left( \frac{d-3}{2},1,1; 1,1,\frac{1}{2}; 
               \frac{d}{2},\frac{d}{2},\frac{d}{2};
               \frac{m_2^2}{R_4},
               \frac{m_2^2}{m_2^2-R_{231}}, 
               \frac{m_2^2}{m_2^2-R_{23} }                        
               \right)  
     +  (1\leftrightarrow 2)     
\Bigg]  \Bigg\}
\nonumber\\
&& 
+\Big\{(2,3,1)\leftrightarrow (3,1,2)
\Big\}. \nonumber
\end{eqnarray}
This representation is valid
if the amplitude of arguments of these 
hypergeometric functions are less than 
$1$ and $\mathcal{R}$e$(d-3)>0$.
\subsection{$R_{ij} = m_i^2$  
or $m_j^2$ for $i,j=1,2,3,4$} 
Taking $R_{12} = m_1^2$ as 
an example, 
one confirms that
\begin{eqnarray}
\partial_2 R_{12} =0.
\end{eqnarray}
As a result, the terms $(1,2,3)$ 
of $J_{1234}^{(d)}$ that are 
multiplied by $\frac{\partial R_2}{\partial m_2^2}$ 
vanish. 
Other terms of $J_{1234}^{(d)}$ are given by
(\ref{j1234normal}). The terms  
$J_{2341}^{(d)}, J_{3412}^{(d)}$
and $J_{4123}^{(d)}$ of $J_4$ are given by 
(\ref{j1234normal}).
\subsection{$G_3 = 0$} 
In this case, one has 
\begin{eqnarray}
J_4
= - \frac{1}{2}\sum\limits_{k=1}^4 
\left(\frac{\partial_k Y_4 }
{Y_4}\right)  {\bf k}^{-}  
J_4(d;\{p_i^2, s, t\}, \{m_i^2\}).
\end{eqnarray}
This equation is equivalent with ($65$) 
in Ref.~\cite{Devaraj:1997es}. The term ${\bf k}^{-}  
J_4(d;\{p_i^2, s, t\}, \{m_i^2\})$ is given by 
$J_3$ in (\ref{J3normal}).
\subsection{$G_{2(ijk)} = 0$  
for $i,j,l =1,2,3,4$}           
In the same notation, $G_{2(ijk)}$ are
the Gram determinants of $J_3$ 
that are obtained
by shrinking $k$-th propagator 
in $J_4$. We take $|G_{2(123)}| = 0$ as 
an example. By using (\ref{g123J30}) for 
$J_{123}^{(d)}$, we then evaluate 
$J_{1234}^{(d)}$ again, the result is
\begin{eqnarray}
J_{1234}^{(d)}
&=&-\dfrac{\sqrt{\pi} 
\Gamma\left(\frac{d-2}{2}\right) } {8\; 
\Gamma(\frac{d-1}{2}) }
\left( \frac{\partial_4 R_4 }{R_4}\right)
\left(
\frac{\partial_3 Y_{123} }{Y_{123} }\right)
\times \nonumber\\
&&\hspace{0cm}\times  
\left[\dfrac{\partial_2 R_{12} }
{\sqrt{1- m_1^2/R_{12} }}
+\dfrac{\partial_1 R_{12} }
{\sqrt{1- m_2^2/R_{12} }}  \right]
\left(R_{12}\right)^{\frac{d-4}{2}} 
\Fh21\Fz{1, \frac{d-3}{2} }{\frac{d-1}{2} }
{\frac{R_{12} }{R_4 } } 
\\
&&+ \dfrac{\Gamma\left(\frac{d-2}{2}\right) } 
{8\;\Gamma(\frac{d}{2}) }
\left( \frac{\partial_4 R_4 }{R_4} \right)
\left(\frac{\partial_3 
Y_{123} }{Y_{123} } \right)
\times 
\nonumber\\
&&  \hspace{0cm}   \times  
 \left[
\dfrac{\partial_2 R_{12} }
{\sqrt{1- m_1^2/R_{12} }}\;
(m_1^2)^{\frac{d-4}{2}}
F_1 \left(\frac{d-2}{2}; 1, \frac{1}{2}; 
\frac{d}{2}; \frac{m_1^2}{R_4}, 
\frac{m_1^2}{R_{12} } \right)
+ ( 1 \leftrightarrow 2)  
\right] \nonumber\\
&& \nonumber\\
&&\hspace{-0cm} 
+
\{(1,2,3)\leftrightarrow (2,3,1)\} 
\quad
+
\quad
\{(1,2,3)\leftrightarrow (3,1,2)
\}. 
\nonumber
\end{eqnarray}
This representation is valid
if the amplitude of arguments of these 
hypergeometric functions
are less than $1$ and 
$\mathcal{R}$e$(d-3)>0$.
\subsection{$G_{1(ij)}= 0$ 
for $i,j=1,2,3,4$}            
One assumes that the term $J_{1234}^{(d)}$ has 
$G_{1(12)}=0$.
Recalculating this term, 
the result reads in term of 
Gauss and Appell $F_3$ functions 
\begin{eqnarray}
J_{1234}^{(d)} & = & 
-\left( \frac{\partial_4 R_4 }{2\; R_4}
\right) \; J_{123}^{(d=4)} 
\; (R_{123})^{ \frac{d-4}{2}}\; 
\Fh21\Fz{1, \frac{d-3}{2} }{\frac{d-2}{2} }
{\frac{R_{123}}{R_4 }}     \\
&&
+ \frac{\Gamma\left(\frac{d-2}{2} \right)}
{4\Gamma\left(\frac{d}{2} \right) } 
\left(\frac{\partial_4 R_4 }{R_4} \right)
\left(
\frac{\partial_3 R_{123} }{R_{123}} 
\right)
\times \nonumber\\
&&\times \Bigg[
\dfrac{R_{123}}{(R_{123} -m_1^2) }  
\dfrac{(m_1^2)^{\frac{d-2}{2}} }{(m_1^2-m_2^2)} 
\; F_3\left(\frac{d-3}{2}, 1; 1, 1; \frac{d}{2}; 
\dfrac{m_1^2}{R_4},
\dfrac{m_1^2}{m_1^2 - R_{123}} \right)
+ (1\leftrightarrow 2) \Bigg]  \nonumber\\ 
&& \nonumber\\
&&\hspace{-0cm} 
+\Bigg\{ 
-\left(\frac{\partial_4 R_4 }
{2 R_4}\right) \; J_{231}^{(d=4)}
\;  \left( R_{231}\right)^{\frac{d-4}{2}} 
\Fh21\Fz{ \frac{d-3}{2}, 1}{\frac{d-2}{2} }
{\frac{R_{231}}{R_4} }
\nonumber\\
&& \hspace{0.7cm}+ \dfrac{\sqrt{\pi} 
\Gamma\left(\frac{d-2}{2}\right) }
{\Gamma(\frac{d-1}{2}) }
\left(
\frac{\partial_4 R_4 }{R_4} 
\right)
\left(
\frac{\partial_1 R_{231} }{R_{231}}
\right)
\left[\dfrac{\partial_3 R_{23} }
{\sqrt{1-m_2^2/R_{23} }}
+ \dfrac{\partial_1 R_{23} }
{\sqrt{1-m_2^2/R_{23} }} \right] 
\times \nonumber\\
&&   \hspace{4cm}          
 \times
 \frac{~~~\left(R_{23}\right)^{\frac{d-4}{2}}}
 {\sqrt{1-R_{23}/R_{231}}}
 F_1\left(\frac{d-3}{2}; 1, \frac{1}{2}; \frac{d-1}{2}; 
  \frac{R_{23} }{R_4}, 
  \frac{R_{23} }{R_{231}}  \right) 
\nonumber\\
&&
\hspace{0.7cm}
-\dfrac{\Gamma\left(\frac{d-2}{2}\right) }
{8\;\Gamma(\frac{d}{2}) }
\left(
\frac{\partial_4 R_4 }{R_4} 
\right)\Bigg[
\frac{\partial_1 R_{231} }{(R_{231} -m_2^2)} 
\frac{\partial_3 R_{23} }
{(R_{23} -m_2^2)} (m_2^2)^{\frac{d-2}{2}}
\times \nonumber\\
&&   \hspace{0.8cm} \times 
F_S\left( \frac{d-3}{2},1,1; 1,1,\frac{1}{2}; 
               \frac{d}{2},\frac{d}{2},\frac{d}{2};
               \frac{m_2^2}{R_4},
               \frac{m_2^2}{m_2^2-R_{231}}, 
               \frac{m_2^2}{m_2^2-R_{23} }                        
               \right)  
     +  (1\leftrightarrow 2)     
\Bigg]  \Bigg\}
\nonumber\\
&& 
+\Big\{(2,3,1)\leftrightarrow (3,1,2)
\Big\}. \nonumber
\end{eqnarray}
Where the terms $J_{123}^{(d=4)}$
and $J_{231}^{(d=4)}$ are given
in (\ref{J123normal4}).
This representation is valid if 
the amplitude of 
arguments of these hypergeometric 
functions are less 
than $1$ and $\mathcal{R}$e$(d-3)>0$. 
The Appell $F_3$
functions are described in detail in 
appendix $B$ (see
(\ref{F3}) in more detail).

For future prospect of this work,  
a package which provides a general
$\epsilon$-expansion and numerical evaluations
for one-loop functions at general $d$ 
is planned. 
To achieve this purpose, many related 
works are worth mentioning
in this paragraph. 
First, automatized analytic continuation 
of Mellin-Barnes integrals have been
presented in \cite{Czakon:2005rk}. 
The construction 
of Mellin-Barnes representations for 
Feynman integrals has been performed in 
\cite{Gluza:2007rt, Gluza:2010rn}. 
Recent development for treating numerically 
Mellin-Barnes integrals in physical regions 
has been
proposed in \cite{Dubovyk:2017cqw,
Usovitsch:2018shx,
Dubovyk:2019krd}. The hypergeometric functions
in this work can be expressed as the 
multi-fold MB integrals and they may be evaluated 
numerically by following the above works. 
Furthermore, 
the $\epsilon$-expansion of the 
hypergeometric functions
appearing in our 
analytic results may be also 
performed by using the packages 
{\tt Sigma}, {\tt EvaluateMultiSums} and 
{\tt Harmonic Sums} \cite{PACKAGES}. 
Numerical $\epsilon$-expansion of hypergeometric 
functions may be done by using 
{\tt NumEXP}~\cite{Huang:2012qz}.
Besides that, analytic  $\epsilon$-expansion
for the hypergeometric functions has 
been carried out in
\cite{Huber:2005yg,Huber:2007dx,Kalmykov:2006hu,
Kalmykov:2006pu,Moch:2005uc,Friot:2011ic,
Greynat:2013hox,Greynat:2013zqa}.  
Differential reduction of generalized 
hypergeometric 
functions has been also reported 
in~\cite{Bytev:2009kb,Bytev:2011ks,
Bytev:2013bqa,Bytev:2013gva}.

In the context 
of dimensional recurrence relations,
the tensor reductions for 
one-loop up to five-point functions 
have been worked out in 
\cite{Fleischer:2010sq} and for
higher-point 
functions have been developed 
in~\cite{Fleischer:2011hc}.
In practice, one encounters integrals
with denominator powers higher than one
and their reduction needs to be considered,
see e.g.~\cite{Davydychev:1990cq} 
for the scalar case.
IBP reduction can be combined with 
dimensional recurrence relations to
reduce them to master integrals of 
higher space-time dimensions.
\section{Conclusions}
In this article, we have been presented the 
analytic results for scalar one-loop 
two-, three- and four-point functions
in detail. The results have been 
expressed in terms 
of Gauss $_2F_1$, Appell {\tt $F_1$} and 
$F_S$ hypergeometric functions. 
All cases of external momentum 
and internal mass assignments have 
considered in detail in this work.
The higher-terms in the $\epsilon$-expansion 
for one-loop integrals can be performed
directly from analytic expressions in this 
work. These terms are necessary building blocks
in computing
two-loop and higher-loop 
corrections. Moreover, one-loop functions in 
arbitrary $d$ in this work may be taken account
in the evaluations for higher-loop Feynman 
integrals. The one-loop functions with 
$d\geqslant 4$ can
also used in the reduction 
for tensor one-loop Feynman integrals. 
For future works,
a package for numerical evaluations 
for one-loop integrals at general $d$ and general
$\epsilon$-expansion for these integrals 
is planned. Additionally, the method 
can extend to evaluate two- and higher-loop 
Feynman integrals. \\

\noindent
{\bf Acknowledgment:}~
This research is funded by Vietnam 
National Foundation for Science and 
Technology Development (NAFOSTED) 
under the grant number 
$103.01$-$2019.346$. The author 
would like to thank T. Riemann 
for helpful discussions and comments. 
\section*{Appendix A: Useful formulas}%
In this appendix, we show
some useful formulas used in this paper. 
We applied the reflect 
formula for gamma functions~\cite{carlson}:
\begin{eqnarray}
\label{reflect}
\Gamma(1-z-n) = (-1)^n
\dfrac{\Gamma(z)\Gamma(1-z)}
{\Gamma(z+n)},
\end{eqnarray}
provided that $z\in \mathbb{C}$ 
and $n \in \mathbb{N}$. We have mentioned 
the duplication 
formula for gamma functions 
~\cite{carlson}:
\begin{eqnarray}
\label{duplication}
\Gamma(2z) = \dfrac{2^{2z-1} \Gamma(z)
\Gamma(z+\frac{1}{2})}{\sqrt{\pi}},
\end{eqnarray}
provided that $z\in \mathbb{C}$.
\section*{Appendix B: 
Generalized hypergeometric series}
Generalized hypergeometric functions
are presented in this appendix.
\subsection*{Gauss hypergeometric 
series}
Gauss hypergeometric series are 
given~\cite{Slater}:
\begin{eqnarray}
\label{gauss-series}
\Fh21\Fz{a,b}{c}{z} 
= \sum\limits_{n=0}^{\infty} 
\frac{(a)_n (b)_n}{(c)_n} 
\frac{z^n}{n!},
\end{eqnarray}
provided that $|z|<1$. 
The pochhammer symbol 
$(a)_n =\Gamma(a+n)/\Gamma(a)$ 
is used.

The integral representation 
for Gauss hypergeometric 
functions~\cite{Slater}
reads
\begin{eqnarray}
\label{gauss-int}
 \Fh21\Fz{a,b}{c}{z} 
= \dfrac{\Gamma(c)}
{\Gamma(b)\Gamma(c-b)} 
\int\limits_0^1 du
\; u^{b-1} (1-u)^{c-b-1} (1-zu)^{-a},
\end{eqnarray}
provided that  $|z|<1$ 
and Re$(c)>$Re$(b)>0$.

Basic linear transformation formulas
for Gauss $_2F_1$ hypergeometric 
functions
which are collected from 
Ref.~\cite{Slater,Barnes}, 
are listed as 
follows:
\begin{eqnarray}
\Fh21\Fz{a,b}{c}{z} &=&
\Fh21\Fz{b,a}{c}{z}  \label{f21a} \\
&=& (1-z)^{c-a-b} 
\Fh21\Fz{c-a,c-b}{c}{z} \label{f21b} \\
&=& (1-z)^{-a} 
\Fh21\Fz{ a,c-b}{c}{\frac{z}{z-1} }
\label{f21c}  \\
&=& (1-z)^{-b} 
\Fh21\Fz{ b,c-a}{c}{\frac{z}{z-1} }
\label{f21d} \\
&=& \frac{\Gamma(c) \Gamma(c-a-b)}
{\Gamma(c-a) \Gamma(c-b)} 
\Fh21\Fz{a,b}{a+b-c+1}{1-z}
\label{f21f} \nonumber \\
&+& (1-z)^{c-a-b} 
\frac{\Gamma(c) \Gamma(a+b-c)}
{\Gamma(a) \Gamma(b)} 
\Fh21\Fz{c-a,c-b}{c-a-b+1}{1-z}\\
&=& \frac{\Gamma(c) \Gamma(b-a)}
{\Gamma(b) \Gamma(c-a)} (-z)^{-a} 
\Fh21\Fz{a,1-c+a}{1-b+a}{\frac{1}{z} }
\label{f21g} \nonumber \\
&+& \frac{\Gamma(c) \Gamma(a-b)}
{\Gamma(a) \Gamma(c-b)} (-z)^{-b} 
\Fh21\Fz{b,1-c+b}{1-a+b}
{\frac{1}{z} }\label{z-}.
\end{eqnarray}
\subsection*{Appell series}
Appell $F_1$ series are 
defined~\cite{Slater,F1Kampe}:
\begin{eqnarray}
\label{appell-series}
F_1(a; b, b'; c; x, y) 
=\sum\limits_{m=0}^{\infty}
\sum\limits_{n=0}^{\infty}
\dfrac{(a)_{m+n} (b)_m (b')_n}
{(c)_{m+n}\; m! n!} 
x^m y^n,
\end{eqnarray}
provided that $|x|<1$ and $|y|<1$. 

The single integral 
representation for $F_1$ reads~\cite{Slater}
\begin{eqnarray}
\label{appell-int}
F_1(a; b, b'; c; x, y) 
= \frac{\Gamma(c)}{\Gamma(c-a)\Gamma(a)} 
\int\limits_0^1 du\;
u^{a-1} (1-u)^{c-a-1} (1-xu)^{-b}(1-yu)^{-b'},
\end{eqnarray}
provided that Re$(c)$ $>$ Re$(a)>0$ and $|x|<1$,
$|y|<1$.

We collect all transformations for Appell 
$F_1$ functions from 
Refs.~\cite{Slater,schlosser}.
The first relation for $F_1$ is mentioned, 
\begin{eqnarray}
\label{firstF1}
F_1\Big(a;b,b';c;x,y\Big)=(1-x)^{-b}(1-y)^{-b'}
F_1\Big(c-a;b,b';c;\frac x{x-1},\frac y{y-1}\Big).
\end{eqnarray}
If $b'=0$, we arrive at the well-known 
Pfaff--Kummer transformation for the $_2F_1$. 
Further, we have
\begin{eqnarray}
F_1\Big(a;b,b';c;x,y\Big)=(1-x)^{-a}
F_1\!\left(a;-b-b'+c,b';c;\frac x{x-1},
\frac{y-x}{1-x}\right).
\end{eqnarray}
Furthermore, if $c=b+b'$, one then obtains
\begin{eqnarray}
F_1\Big(a;b,b';b+b';x,y\big)&=&(1-x)^{-a}
\Fh21\Fz{a,b'}{b+b'}{\frac{y-x}{1-x}}\label{cbb1}\\
&=&(1-y)^{-a}\Fh21\Fz{a,b}{b+b'}{\frac{x-y}{1-y}}.
\label{cbb2}
\end{eqnarray}
Similarly,
\begin{eqnarray}
 F_1\left(a;b,b';c;x,y\right)=(1-y)^{-a}
 F_1\!\left(a;b,c-b-b';c;\frac{x-y}{1-y},
 \frac y{y-1}\right),
 \end{eqnarray}
and 
\begin{eqnarray}
 F_1\left(a;b,b';c;x,y\right)
 ={\mbox{\small $(1-x)^{c-a-b}(1-y)^{-b'}$}}
 F_1\!\left(c-a;c-b-b',b';c;x,\frac{x-y}{1-y}\right),\\
 F_1\left(a;b,b';c;x,y\right)
 ={\mbox{\small $(1-x)^{-b}(1-y)^{c-a-b'}$}}
 F_1\!\left(c-a;b,c-b-b';c;\frac{y-x}{1-x},y\right).
 \end{eqnarray}
A further relation for $F_1$ is given 
\begin{eqnarray}
\label{f1relation}
F_1\Big(a+1, 1, 1/2, a+2; x, y\Big) 
&=&  \dfrac{\sqrt{\pi}\Gamma(a+2)}{\Gamma(a+3/2)}\;
y^{-a-1}\;\Fh21\Fz{1, a+1}{a+3/2 }{\dfrac{x}{y}   }\\
& &-\frac{2\Gamma(a+2)}{\;\Gamma(a+1)}
\dfrac{\sqrt{1-y}}{y(1-x)} \;
F_1\Bigg(1, -a, 1, 3/2; 1-\dfrac{1}{y} ,
\dfrac{x(1-y)}{y(1-x)}\Bigg)
.\nonumber
\end{eqnarray}
\subsection*{Appell $F_3$ series}
Appell $F_3$ series
are written~\cite{Slater,F1Kampe}
\begin{eqnarray}
\label{F3}
F_3\left(a,a';b,b';c;x,y\right)&=&
\sum_{m\ge0}\sum_{n\ge0}
\frac{(a)_m\,(a')_n\,(b)_m\,(b')_n}
{m!\,n!\ (c)_{m+n}}x^my^n,
\end{eqnarray}
provided that Re$(c)$ $>$ Re$(a)>0$ and $|x|<1$, 
$|y|<1$.
\subsection*{ J.~Kamp\'e de F\'eriet series} 
In addition, J.~Kamp\'e de F\'eriet series
\cite{F1Kampe,KampedeFeriet:37}
with two variables are shown:
\begin{eqnarray}
\label{Kamp}
&&\hspace{-2cm}F^{p:q}_{r:s}
\left(\begin{matrix}a_1,
\dots,a_p:b_1,b_1';\dots;b_q,b_q';\\
c_1,\dots,c_r:d_1,d_1';
\dots;d_s,d_s';\end{matrix}\,x,y\right)=\\
&&=
\sum_{m\ge0}\sum_{n\ge0}
\frac{(a_1)_{m+n}\dots(a_p)_{m+n}}
{(c_1)_{m+n}\dots(c_r)_{m+n}}
\frac{(b_1)_m(b_1')_n\dots(b_q)_m(b_q')_n}
{(d_1)_m(d_1')_n\dots(d_s)_m(d_s')_n}
\frac{x^my^n}{m!n!}. \nonumber
\end{eqnarray}
\subsection*{Lauricella-Saran function $F_S$ }
Lauricella-Saran function $F_S$ is defined in terms
of a triple hypergeometric series~\cite{Saran55}
\begin{eqnarray}
\label{lauricella}
&&F_S(\alpha_1,\alpha_2,\alpha_2,
      \beta_1, \beta_2, \beta_3,
      \gamma_1,\gamma_1,\gamma_1; x,y,z) 
   =   
  \sum_{r,m,n=0}^{\infty}
      \frac{(\alpha_1)_r (\alpha_2)_{m+n} (\beta_1)_{r}
      (\beta_2)_m (\beta_3)_n }
      {(\gamma_1)_{r+m+n}~ r! m! n!}
      x^r y^m z^n.\nonumber\\
\end{eqnarray}
The integral representation of $F_S$ 
is defined~\cite{Saran55}:
\begin{eqnarray}
\label{lauricella-int}
&&F_S(\alpha_1,\alpha_2,\alpha_2,
      \beta_1, \beta_2, \beta_3,
      \gamma_1,\gamma_1,\gamma_1; x,y,z) =\\
&&\hspace{2cm}
= \frac{\Gamma (\gamma_1)}{\Gamma (\alpha_1 ) \Gamma (\gamma_1-\alpha_1)}
\int_0^1  dt\; \dfrac{t^{\gamma_1-\alpha_1-1}(1-t)^{\alpha_1-1}}
{(1-x+tx)^{\beta_1}} F_1\left(\alpha_2,\beta_2,\beta_3;
\gamma_1-\alpha_1; ty,tz\right), \nonumber
\end{eqnarray}
provided that Re$(\gamma_1-\alpha_1-\alpha_2)>0$, 
Re$(\alpha_1)>0$, Re$(\alpha_2)>0$, 
and $|x|,\;|y|,\; |z|<1$.
\section*{Appendix C: The contour integrations}
\subsection*{Type 1}
Mellin-Barnes relation~\cite{watson} is given:
\begin{eqnarray}
\label{MB}
\dfrac{1}{2\pi i} 
\int\limits_{-i\infty}^{+i\infty}ds \;
\dfrac{\Gamma(-s)\;\Gamma(\lambda +s)}
{\Gamma(\lambda) } \; \left(z\right)^s
=
\dfrac{1}{(1 + z)^{\lambda}} 
\end{eqnarray}
provided that
$|\mathrm{Arg}(z)|<\pi$. 
The integration contour 
is chosen
in such a way that 
the poles of $\Gamma(- s)$ 
and $\Gamma(\lambda + s)$ are 
well-separated.
\subsection*{Type 2}
The Barnes-type integral for Gauss 
hypergeometric~\cite{Slater} reads 
\begin{eqnarray}
\label{MB2F1}
\dfrac{1}{2\pi i} 
\int\limits_{-i\infty}^{+i\infty}ds \;
\dfrac{\Gamma(-s)\;\Gamma(a +s) 
\;\Gamma(b+s)}{\Gamma(c+s) } 
\left(-z \right)^s
= 
\dfrac{\Gamma(a)\Gamma(b)}{\Gamma(c)}
\Fh21\Fz{a,b}{c}{z}
\end{eqnarray}
provided that $|\mathrm{Arg}(-z)|<\pi$ 
and $|z|<1$. 
\subsection*{Type 3}
The next Barnes-type integral 
applied in this 
paper is
\begin{eqnarray}
\label{MBF1}
\dfrac{1}{2\pi i} 
\int\limits_{-i\infty}^{+i\infty}ds \;
\dfrac{\Gamma(-s)\;\Gamma(a +s) \;
\Gamma(b+s)}{\Gamma(c+s) }
(-x)^s \; \Fh21\Fz{a+s,b'}{c+s}{y}
=
\dfrac{\Gamma(a)\Gamma(b)}{\Gamma(c)}
F_1 (a; b, b';c; x,y)
\nonumber\\
\end{eqnarray}
with $|\mathrm{Arg}(-x)|<\pi$, 
$|\mathrm{Arg}(-y)|<\pi$ and $|x|<1$ and $|y|<1$. 
Under these conditions, 
one could close the 
contour of integration to the right.
Subsequently, we have to take into 
account the residua of the sequence 
poles of $\Gamma(-s)$.
The result is expressed
as the summation of Gauss hypergeometric
function. The summation is then identical as
a series of Appell $F_1$ functions
~\cite{schlosser} as follows:
\begin{eqnarray}
\label{MBF1-a}
\dfrac{\Gamma(a)\Gamma(b)}{\Gamma(c)}
\sum\limits_{m=0}^{\infty}
\dfrac{(a)_m (b)_m}{(c)_m} 
\dfrac{x^m}{m!} \Fh21\Fz{a+m,b'}{c+m}{y}
= \dfrac{\Gamma(a)\Gamma(b)}{\Gamma(c)}
F_1 (a; b, b';c; x,y).
\end{eqnarray}
\subsection*{Type 4}
Furthermore, in this paper we evaluate 
the following integral:
\begin{eqnarray}
\label{MBFS}
&&\dfrac{1}{2\pi i} 
\int\limits_{-i\infty}^{+i\infty}ds \;
\dfrac{\Gamma(-s)\;\Gamma(a +s) \;
\Gamma(b_1+s)}{\Gamma(c+s) }
(-x)^s \; F_1 (a' + s; b_2, b_3; c+s; y,z)
\\
&&=
\dfrac{\Gamma(a)
\Gamma(b_1)}{\Gamma(c)}(1-y)^{-b_2} (1-z)^{-b_3} 
\; F_S \left(a, c-a', c-a'; b_1, b_2, b_3; c,c,c;
x, \frac{y}{y-1}, \frac{z}{z-1} \right).
\nonumber
\end{eqnarray}
We close the contour of integration to
the right. By taking into account 
the residua of the sequence 
poles of $\Gamma(-s)$, 
the result reads
\begin{eqnarray}
\label{fsf1}
\dfrac{\Gamma(a)\Gamma(b_1)}{\Gamma(c)}
\sum\limits_{m=0}^{\infty}
\dfrac{(a)_m (b_1)_m}{(c)_m} 
\dfrac{x^m}{m!}\; F_1(a'+m; b_2, b_3; c+m; y, z). 
\end{eqnarray}
One applies the relation~(\ref{firstF1})
for Appell $F_1$ functions in 
(\ref{fsf1}) as follows
\begin{eqnarray}
F_1(a'+m; b_2, b_3; c+m; y, z)
= (1-y)^{-b_2} (1-z)^{-b_3} 
F_1 \left(c-a'; b_2, b_3;c+m; \frac{y}{y-1},
\frac{z}{z-1} \right). \nonumber\\
\end{eqnarray}
Eq.~(\ref{fsf1}) is then presented as a series of 
Lauricella functions $F_S$~\cite{Saran55}
\begin{eqnarray}
\label{MBFS-b}
&&\hspace{0cm}\dfrac{\Gamma(a)\Gamma(b_1)}
{\Gamma(c)}(1-y)^{-b_2} (1-z)^{-b_3} 
\times \\
&& 
\times \sum\limits_{m, n, l=0}^{\infty}
\dfrac{(a)_m (b_1)_m (b_2)_n (b_3)_l
(c-a')_{n+l}}{(c)_{m+n+l}} 
\dfrac{x^m}{m!}\; \frac{1}{n!} 
\left(\frac{y}{y-1} \right)^n \frac{1}{l!}
\left( \frac{z}{z-1}) \right)^l =                                 
\nonumber\\
&& =\dfrac{\Gamma(a)
\Gamma(b_1)}{\Gamma(c)}(1-y)^{-b_2} 
(1-z)^{-b_3} 
\; F_S \left(a, c-a', c-a'; b_1, b_2,
b_3; c,c,c;
x, \frac{y}{y-1}, \frac{z}{z-1} \right).
\nonumber
\end{eqnarray}
provided that $|x|<1$, 
$\left|\frac{y}{y-1}\right|<1$
and $\left|\frac{z}{z-1}\right|<1$. 

The last formula is mentioned in this paper relates to a
transformation of Mellin-Barnes integrals (see page $156$ 
of Ref.~\cite{Barnes}
, item $14.53$ in page $290$ of~\cite{watson}) which is
\begin{eqnarray}
\label{MB-transform}
&&\hspace{-1cm} \int\limits_{-i\infty}^{+i\infty}ds \;
  \dfrac{\Gamma(-s)\;\Gamma(a +s) \;\Gamma(b+s)}{\Gamma(c+s) } (-z)^s =   \\
&&\hspace{-0.5cm}= \int\limits_{-i\infty}^{+i\infty}ds \; 
  \dfrac{ \Gamma(-s)\; \Gamma(a +b-c - s) \Gamma(c-a + s) \Gamma(c-b + s) }
        {\Gamma(c-a)\Gamma(c-b) }\; (1-z)^{c-a-b+s}, \nonumber
\end{eqnarray}
provided that $|\mathrm{Arg}(-z)|<2\pi$. 
\section*{Appendix $D$: Master equation 
for $J_N$}

General relation for $J_N$ have 
been proved in \cite{Bluemlein:2017rbi}. 
In this appendix, we consider $N=4$
as an example. Performing Feynman 
parameterization for $J_4$, one 
arrives at
\begin{eqnarray}
\label{feyn-d0}          
\dfrac{J_4}{\Gamma\Big(4-\frac{d}{2}\Big)}
&=&
\int dS_3 \Big[ 
a x_1^2 + bx_2^2 + cx_3^2 + 2\;d\; x_1 x_2 
+ 2\; e\; x_1 x_3 + 2\; f\; x_2 x_3 + g x_1 
+ hx_2 +k x_3 +j -i\rho
\Big]^{\frac{d}{2} -4}.
\nonumber\\
\end{eqnarray}
Above coefficients $a, b, c,\cdots,j$ 
are shown \\

\begin{tabular}{l@{\hspace{1.5cm}}l@{\hspace{1.5cm}}l}
$a = p_1^2,$ & $b = p_2^2,$ & $c = (p_2+p_3)^2,$ \\[2ex]
$d = -p_1p_2,$ & $e = - p_1 p_2- p_1p_3,$ & 
$f = p_2^2 + p_2p_3,$ \\[2ex]
$g = m_1^2-m_2^2-p_1^2, $ & $h = -m_2^2+m_3^2-p_2^2,$ 
&$k = -m_2^2+m_4^2 -(p_2+p_3)^2,$ \\[2ex]
$j=m_2^2$. &  &   \\ 
\end{tabular}
\\

We used the following notation
\begin{eqnarray}
\int \;dS_3 &=&
\int\limits_0^1 dx_1\int\limits_0^{1-x_1}dx_2
\int\limits_0^{1-x_1-x_2} dx_3     
=  
\int\limits_0^1 dx_2\int\limits_0^{1-x_2}dx_3
\int\limits_0^{1-x_2-x_3} dx_1 
=
\int\limits_0^1 dx_3\int\limits_0^{1-x_3}dx_1
\int\limits_0^{1-x_1-x_3} dx_2. \nonumber\\
\label{ds3}
\end{eqnarray}
The integrand of $J_4$ is 
\begin{eqnarray}
\mathcal{M}_4(x_1,x_2,x_3) 
&=&a x_1^2 + bx_2^2 + cx_3^2 
+ 2\;d\; x_1 x_2 + 2\; e\; x_1 x_3 + 2\; f\; x_2 x_3 
+ g x_1 + hx_2 +k x_3 +j. 
\nonumber\\
&=&(x_1,x_2, x_3) \; \mathcal{G}_3 
\left(
\begin{array}{c}
x_1\\
x_2\\
x_3\\
\end{array}
\right)
+2 \mathcal{H}^T_4 \left(
\begin{array}{c}
x_1\\
x_2\\
x_3\\
\end{array}
\right)
+\mathcal{K}_4                                \\
&=& (x_1 -y_1,x_2-y_2, x_3-y_3)
\; \mathcal{G}_3
\left(
\begin{array}{c}
x_1-y_1\\
x_2-y_2\\
x_3-y_3\\
\end{array}
\right)
+R_4 \\
&=& 
\Lambda_4(x_1,x_2, x_3) +R_4.
\end{eqnarray}
These matrices are given
\begin{eqnarray}
\mathcal{G}_3 
=  \left( \begin{array}{ccc}
            a  &  d & e \\
            d  &  b & f \\
            e  &  f & c 
\end{array} \right),
\quad 
\mathcal{H}_4 =\frac{1}{2}\left( \begin{array}{c}
                         g  \\
                         h  \\
                         k  \\
\end{array} \right), 
\quad \mathcal{K}_4 = j. 
\nonumber
\end{eqnarray}
The $\Lambda_4(x_1,x_2, x_3)$ reads
\begin{eqnarray}
\Lambda_4(x_1,x_2, x_3) &=&  
      a(x_1-y_1)^2 + b(x_2-y_2)^2 + c(x_3-y_3)^2
    + 2d (x_1-y_1)(x_2-y_2) \nonumber\\
&&  + 2e (x_1-y_1)(x_3-y_3) + 2f(x_2-y_2)(x_3-y_3). 
\end{eqnarray}
The vector $\overrightarrow{y}$ is defined as: 
$\overrightarrow{y}= (y_1, y_2, y_3)  
= - \mathcal{G}_3^{-1} H^T$. 
We write explicitly $y_1, y_2, y_3$
and $y_4$ as follows:
\begin{eqnarray}
y_1 &=& \dfrac{-b c h+b e k+c d g-d f k-e f g+f^2 h}
{ G_3} 
=\dfrac{\partial R_4}{\partial m_1^2}, \nonumber\\
y_2 &=& \dfrac{a (f k-c g)+c d h-e (d k+f h)+e^2 g}{ G_3}
=\dfrac{\partial R_4}{\partial m_3^2}, \\
y_3 &=& \dfrac{a f g + b e h - d (e g + f h) - a b k + d^2 k}{ G_3}
=\dfrac{\partial R_4}{\partial m_4^2},\nonumber\\
y_4 &=& 1- y_1-y_2-y_3 
= \dfrac{\partial R_4}{\partial m_2^2}. \nonumber
\end{eqnarray}
First, we consider $G_3 \neq 0$ and 
$R_4 \neq 0$. In this case, 
Mellin-Barnes relation is applied 
to decompose $J_4$'s integrand as follows: 
\begin{eqnarray}
&&\hspace{-0.5cm}
\dfrac{1}{\Big[ \Lambda_4(x_1,x_2, x_3)
+ R_4-i\rho\Big]^{4-\frac{d}{2}}}
=\nonumber\\
&&\hspace{1cm}=
\dfrac{1}{2\pi i} \int\limits_{-i\infty}^{+i\infty}ds \; 
\dfrac{\Gamma(-s)\;
\Gamma( 4-\frac{d}{2} +s)}
{ \Gamma(4-\frac{d}{2} ) }    
\left(\dfrac{1}{R_4-i\rho} 
\right)^{4-\frac{d}{2} }
\; 
\left[\dfrac{\Lambda_4(x_1,x_2,x_3)}{ R_4-i\rho} 
\right]^s, 
\end{eqnarray}
provided that 
$\left|\mathrm{Arg}\left(\frac{\Lambda_4(x_1,x_2,x_3)}
{R_4-i\rho} \right)\right|<\pi$. 
With the help of the Mellin-Barnes relation, 
this brings the Feynman parameters integration
to the simpler form:
\begin{eqnarray}
 \mathcal{F}_4 = \int dS_3\; 
\left[\dfrac{\Lambda_4(x_1,x_2, x_3)}
{ R_4-i\rho} \right]^s.
\end{eqnarray}
In order to carry out this integral, 
we consider the following differential operator
(see theorem of Bernshtein
\cite{Bernshtein}, or \cite{Golubeva})
\begin{eqnarray}
\hat{O}_4 = \frac{1}{2} (x_1-y_1)\frac{\partial}{\partial x_1}
         + \frac{1}{2} (x_2-y_2)\frac{\partial }{\partial x_2}  
         + \frac{1}{2} (x_3-y_3)\frac{\partial }{\partial x_3}.
\end{eqnarray}
It is easy to check that 
\begin{eqnarray}
\hat{O}_4\;  
\left[\dfrac{\Lambda_4(x_1,x_2, x_3)}{ R_4 } \right]^s 
= s \;\left[\dfrac{\Lambda_4(x_1,x_2,x_3)}{ R_4} \right]^s.
\end{eqnarray}
As a matter of this fact, 
we can rewrite Feynman parameter 
integral as
\begin{eqnarray}
\mathcal{F}_4 
&=& \dfrac{1}{s} \int dS_3\;  
    \hat{O}_4 
    \left[ \dfrac{\Lambda_4(x_1,x_2, x_3)}{R_4 } \right]^s 
=   \nonumber\\
&=& \dfrac{1}{s} \Bigg\{ 
    \int\limits_0^1 dx_1\int\limits_0^{1-x_1}dx_2
    \int\limits_0^{1-x_1-x_2} dx_3\; (x-y_3)
    \frac{\partial}{ \partial x_3}
    \left[ \dfrac{\Lambda_4(x_1,x_2, x_3)}{R_4 } \right]^s
    \nonumber\\
&&  \hspace{0.5cm}+    
    \int\limits_0^1 dx_1\int\limits_0^{1-x_1}dx_3
    \int\limits_0^{1-x_1-x_3} dx_2\;(x-y_2)
    \frac{\partial}{ \partial x_2}  
    \left[ \dfrac{\Lambda_4(x_1,x_2, x_3)}{R_4 } \right]^s
    \nonumber\\
&& \hspace{0.5cm}+
   \int\limits_0^1 dx_2\int\limits_0^{1-x_3}dx_3 
   \int\limits_0^{1-x_2-x_3} dx_1 \;(x-y_1)
   \frac{\partial}{ \partial x_1} \;\;\; 
   \left[ \dfrac{\Lambda_4(x_1,x_2, x_3)}{R_4 } \right]^s
   \Bigg\}\\
&=& \dfrac{1}{2s} \Bigg\{ 
    \int\limits_0^1 dx_1\int\limits_0^{1-x_1}dx_2
    \int\limits_0^{1-x_1-x_2} dx_3\; 
    \frac{\partial}{ \partial x_3}
    \Big\{ (x_3-y_3)
    \left[ \dfrac{\Lambda_4(x_1,x_2, x_3)}{R_4 } \right]^s
    \Big\}
    \nonumber\\
&&  \hspace{0.5cm}+ 
    \int\limits_0^1 dx_1\int\limits_0^{1-x_1}dx_3
    \int\limits_0^{1-x_1-x_3} dx_2\;
    \frac{\partial}{ \partial x_2}  
    \Big\{
    (x_2-y_2) \left[ \dfrac{\Lambda_4(x_1,x_2, x_3)}{R_4 } \right]^s
    \Big\}
    \nonumber\\
&&  \hspace{0.5cm}+   
    \int\limits_0^1 dx_2\int\limits_0^{1-x_3}dx_3
    \int\limits_0^{1-x_2-x_3} dx_1 \;
    \frac{\partial}{ \partial x_1} 
    \Big\{
    (x_1-y_1)
    \left[ \dfrac{\Lambda_4(x_1,x_2, x_3)}{R_4 } \right]^s
    \Big\}
    \;\;\; \Bigg\} \nonumber\\
&&-\dfrac{3}{2s} \int dS_3 
   \left[ \dfrac{\Lambda_4(x_1,x_2, x_3)}{R_4 } \right]^s.
\label{j41}
\end{eqnarray}
The last term in this equation is proportional to 
$\mathcal{F}_4$. 
It is then combined with $\mathcal{F}_4 $ on the 
left side of Eq. ~(\ref{j41}).
As a result, Eq.~(\ref{j41}) is then casted 
into the form:
\begin{eqnarray}
\label{j42}
\mathcal{F}_4  
&=& \dfrac{\Gamma( s+\frac{3}{2} )}
{2\;\Gamma( s + \frac{5}{2} )} 
    \Bigg\{ \;\;
    \int\limits_0^1 dx_1\int\limits_0^{1-x_1}dx_2
    \int\limits_0^{1-x_1-x_2} dx_3\; 
    \frac{\partial}{ \partial x_3} 
    \Big\{
     (x_3-y_3) \left[ \dfrac{\Lambda_4(x_1,x_2, x_3)}{R_4 } \right]^s
    \Big\}
    +\nonumber \\
& & \hspace{2.5cm} + \int\limits_0^1 
dx_1\int\limits_0^{1-x_1}dx_3
    \int\limits_0^{1-x_1-x_3} dx_2\;
    \frac{\partial}{ \partial x_2} 
    \Big\{
     (x_2-y_2) 
     \left[\dfrac{\Lambda_4(x_1,x_2, x_3)}{R_4 } \right]^s
    \Big\} \\
& & \hspace{2.5cm}+  \int\limits_0^1 dx_2
    \int\limits_0^{1-x_3}dx_3\int\limits_0^{1-x_2-x_3} dx_1 \;
    \frac{\partial}{ \partial x_1} 
    \Big\{
     (x_1-y_1) \left[ \dfrac{\Lambda_4(x_1,x_2, x_3)}{R_4 } \right]^s
    \Big\}
    \;\;\Bigg\}. \nonumber
\end{eqnarray}
Taking over a Feynman parameter 
integration in Eq.~(\ref{j42}), 
the result reads
\begin{eqnarray}
\label{j43}
\mathcal{F}_4 
&=& \dfrac{\Gamma( s + \frac{3}{2} )}{2\; \Gamma(s + \frac{5}{2} )}
    \sum\limits_{i=1}^{4}\; y_i \int dS_2 
     \left[ \dfrac{A_i x_1^2 + B_i x_2^2 + 2 C_i x_1 x_2 
           + D_i x_1 + E_i x_2 + F_i}{R_4 } -1
     \right]^s \nonumber\\          
&=& \dfrac{\Gamma(s+\frac{3}{2})}{2\;\Gamma(s+\frac{5}{2})}
     \sum\limits_{i=1}^{4}\;y_i \int dS_2 
     \left[ \dfrac{\mathcal{M}_3(A_i, B_i, C_i, D_i, E_i, F_i)}{R_4  } - 1
     \right]^s.
\end{eqnarray}                  
The coefficients $A_i, B_i, C_i, D_i, E_i, F_i  $ 
are given in Table~(\ref{ABCFD0}).
\begin{table}[H]
\begin{center}
\begin{tabular}{|c@{\hspace{2cm}}|l@{\hspace{2cm}}
|l@{\hspace{2cm}}|l@{\hspace{2cm}}l|}\hline 
$ i$    & $ 1$ & $2$      & $3$  & $4$ \\ \hline
$A_i$   & $b$  &  $a$     & $a$  & $ a+c-2 e$  \\
$B_i$   & $c$  &  $c$     & $b$  & $a+b-2 d $  \\
$C_i$   & $f$  &  $e$     & $d$  & $a-d-e+f $  \\
$D_i$   & $h$  &  $g$     & $g$  & $-2 a + 2 e - g + k $ \\
$E_i$   & $k$  &  $k$     & $h$  & $-2 a+2 d-g+h $  \\
$F_i$   & $j$  &  $j$     & $j$  & $ a + g + j$  \\ \hline
\end{tabular}
\caption{\label{ABCFD0} The coefficients $A_i, B_i, 
C_i, D_i, E_i, F_i$. }
\end{center}
\end{table}
We then write the coefficients in terms of external 
momenta and internal masses which are given in 
Table~(\ref{ABCFD0-p2m2}).
\begin{table}[H]
\begin{center}
\begin{tabular}{|c|l|l|l|l|}\hline 
$ i$    & $ 1$ & $2$        & $3$  & $4$ \\ \hline
$A_i$   & $p_2^2$           &  $p_1^2$                            & $p_1^2$                 & $ p_4^2$  \\
$B_i$   & $t$               &  $t$                                & $p_2^2$                 & $ s$  \\
$C_i$   & $ p_2(p_2+p_3)$   &  $ -p_1(p_2+p_3)$                   & $-p_1p_2$               & $ -p_4(p_1+p_2) $  \\
$D_i$   & $-(p_2^2+m_2^2-m_3^2)$  &  $-(p_1^2+m_2^2-m_1^2)$       & $-(p_1^2-m_1^2+m_2^2)$  & $-(p_4^2+m_1^2-m_4^2) $ \\
$E_i$   & $-(t+m_2^2-m_4^2)$&  $-( t + m_2^2-m_4^2)$ & $-(p_2^2-m_3^2+m_2^2)$               & $-(s+m_1^2-m_3^2) $  \\
$F_i$   & $m_2^2$           &  $m_2^2$                            & $m_2^2$                 & $ m_1^2$  \\ \hline
\end{tabular}
\caption{\label{ABCFD0-p2m2} The coefficients $A_i, B_i, C_i, D_i, E_i, F_i$ in terms of 
external momenta and internal masses. }
\end{center}
\end{table}

Now we can write the Mellin-Barnes integral for $J_4$ as 
\begin{eqnarray}
J_4 (d;\{p_i^2, s, t\},\{m_i^2\})
&=& \dfrac{1}{2\pi i} \int\limits_{-i\infty}^{+i\infty}ds  \; 
\dfrac{\Gamma(-s)\;
\Gamma( 4-\frac{d}{2} +s)
\Gamma(s+\frac{3}{2})}
{ 2\Gamma(s+\frac{5}{2}) }   
\;\left(\dfrac{1}{R_4}
\right)^{\frac{d}{2}-4}                                                  
\\
&&\hspace{1cm} 
\times  \sum\limits_{i=1}^{4}
\left(\frac{\partial R_4 }{\partial m_i^2} \right) \int dS_2 
\Big[ \frac{\mathcal{M}_3
(A_i, B_i, C_i, D_i, E_i, F_i)}{R_4}-1 \Big]^s.                        
\nonumber
\end{eqnarray}
By applying the relation (\ref{MB-transform}), one gets 

\begin{eqnarray}
J_4 (d;\{p_i^2, s, t\},\{m_i^2\})    
&=& \dfrac{1}{2\pi i} \int\limits_{-i\infty}^{+i\infty}ds \; 
    \dfrac{\Gamma(-s)\; \Gamma(\frac{d-3}{2}+s) \Gamma(s+1) }
          { 2\Gamma(\frac{d-3}{2}) } 
    \left(\dfrac{1}{R_4 } 
    \right)^{4-\frac{d}{2} } 
    \times \\
&&  \times \sum\limits_{i=1}^{4} 
    \left(\frac{\partial R_4 }{\partial m_i^2} \right) 
    \Gamma( 3-\frac{d}{2} -s)   
    \int dS_2 
    \Big[ 
         \frac{\mathcal{M}_3
         (A_i, B_i, C_i, D_i, E_i, F_i)}{R_4} 
     \Big]^{\frac{d}{2}-3+s}.                             
     \nonumber
\end{eqnarray}
This equation is then written as follows:
\begin{eqnarray}
\label{J4J3}
J_4 (d;\{p_i^2, s, t\},\{m_i^2\})
&=& -\dfrac{1}{2\pi i} \int\limits_{-i\infty}^{+i\infty}ds \; 
     \dfrac{\Gamma(-s)\; \Gamma(\frac{d-3}{2}+s) \Gamma(s+1) }
           { 2\Gamma(\frac{d-3}{2}) } 
     \left(\frac{1}{R_4}\right)^s                \times\\
&&\hspace{1.5cm}\times     \sum\limits_{k=1}^4 
     \left( \frac{1}{R_4} 
            \frac{\partial R_4 }{\partial m_k^2} 
      \right) \;
     {\bf k}^- J_4 (d+2s;\{p_i^2, s, t\},\{m_i^2\}).  
\nonumber
\end{eqnarray}
In the case of $R_4 = 0$, 
there is no Mellin-Barnes integral 
for $J_4$. We only apply the ring 
operator $\hat{O}_4$, the result 
arrives
\begin{eqnarray}
\label{rijkl0-sol}
J_4(d;\{p_i^2, s, t\}, \{m_i^2\}) =
\frac{1}{d-5} \sum\limits_{k=1}^4
\left(\dfrac{\partial_k Y_4}{G_3} \right) 
{\bf k^{-} } J_4(d-2;\{p_i^2, s, t\}, \{m_i^2\}).
\end{eqnarray}

\end{document}